\documentclass[prl,twocolumn,floatfix,superscriptaddress]
{revtex4}
\usepackage{graphics,epsfig,amsfonts,amssymb,amsmath}
\usepackage{soul} 
\usepackage[dvipsnames]{xcolor}

\newcommand{\beginsupplement}{%
        \setcounter{table}{0}
        \renewcommand{\thetable}{S\arabic{table}}%
        \setcounter{figure}{0}
        \renewcommand{\thefigure}{S\arabic{figure}}%
     }

\begin{document}
\title{Internal screening and dielectric engineering in magic-angle twisted bilayer graphene}
\author{J. M. Pizarro}
\email{jpizarro@uni-bremen.de}
\affiliation{Institute for Theoretical Physics, University of Bremen, Otto-Hahn-Allee 1, 28359 Bremen, Germany}
\affiliation{Bremen Center for Computational Material Sciences, University of Bremen, Am Fallturm 1a, 28359 Bremen, Germany}
\author{M. R\"osner}
\affiliation{Institute for Molecules and Materials, Radbound University, NL-6525 AJ
Nijmegen, The Netherlands}
\author{R. Thomale}
\affiliation{Institute for Theoretical Physics and Astrophysics, University of W\"urzburg, Am Hubland, D-97074 W\"urzburg, Germany} 
\author{R. Valent\'i}
\affiliation{Institute of Theoretical Physics, Goethe University Frankfurt am Main, D-60438 Frankfurt am Main, Germany}
\author{T. O. Wehling}
\email{twehling@uni-bremen.de}
\affiliation{Institute for Theoretical Physics, University of Bremen, Otto-Hahn-Allee 1, 28359 Bremen, Germany}
\affiliation{Bremen Center for Computational Material Sciences, University of Bremen, Am Fallturm 1a, 28359 Bremen, Germany}


%
\date{\today}
\begin{abstract}  
Magic-angle twisted bilayer graphene (MA-tBLG) has appeared as a tunable testing ground to investigate the conspiracy of electronic interactions, band structure, and lattice degrees of freedom to yield exotic quantum many-body ground states in a two-dimensional Dirac material framework. While the impact of external parameters such as doping or magnetic field can be conveniently modified and analyzed, the all-surface nature of the quasi-2D electron gas combined with its intricate internal properties pose a challenging task to characterize the quintessential nature of the different insulating and superconducting states found in experiments. We analyze the interplay of internal screening and dielectric environment on the intrinsic electronic interaction profile of MA-tBLG. We find that interlayer coupling generically enhances the internal screening. The influence of the dielectric environment on the effective interaction strength depends decisively on the electronic state of MA-tBLG. Thus, we propose the experimental tailoring of the dielectric environment, e.g. by varying the capping layer composition and thickness, as a promising pursuit to provide further evidence for resolving the hidden nature of the quantum many-body states in MA-tBLG.
\end{abstract} 
\maketitle

\textit{Introduction.} 
Stacking two graphene layers at a twist angle $\theta$ on top of each other leads to twisted bilayer graphene (tBLG) featuring a moir\'e pattern with an intricate emergent low-energy electronic structure. For small twist angles $\theta < 2 \, ^{\circ}$, the resulting superlattices host several thousand atoms per unit cell. In this situation, the electronic bands around the charge neutrality point (CNP) become very flat \cite{SuaPRB822010, BisPNAS1082011}, which facilitates strong correlation effects. Recent experiments \cite{CaoN5562018ins,CaoN5562018sc,Yankowitz1059,2019arXiv190102997C} reported the emergence of possibly unconventional superconducting and insulating states in magic-angle tBLG (MA-tBLG) at different levels of doping. The insulating states occur for commensurate fillings at both electron and hole dopings \cite{CaoN5562018ins,Yankowitz1059,2019arXiv190306513L}, signaling a possible Mott-Hubbard origin \cite{KosPRX82018,RadPRB982018,Pizarro_2019}. 
Around these insulating states, superconductivity emerges \cite{CaoN5562018sc,Yankowitz1059,2019arXiv190306513L}, resembling the phase diagram of high-$T_c$ cuprate superconductors \cite{KeiN5182015,KorLTP412015} and other unconventional superconductors \cite{KurPRL952005,SipNM72008}. Different models \cite{PhysRevX.8.031089,2018arXiv180311190R,PhysRevB.98.075154,PhysRevLett.121.217001,RadPRB982018,PhysRevB.98.241407,PhysRevX.8.041041,KosPRX82018,Pizarro_2019,PhysRevLett.121.257001,PhysRevB.98.075109,OchPRB982018,Guinea13174,2018arXiv180802482P,LAKSONO201838,VenPRB982018,PhysRevB.98.241412,2018arXiv181004159K,2018arXiv181008642K,2018arXiv181202550S,2018arXiv181204213X} have been proposed to understand the physics behind these insulating and superconducting states. 

Electronic correlations in ultrathin systems such as MA-tBLG depend decisively on the effective electron-electron interaction profile, which is determined by a delicate interplay of screening processes taking place in the dielectric environment and the material itself \cite{WehPRL1062011,RosPRB922015,schonhoff_interplay_2016}. The profile of the effective electron-electron interactions presents a central uncertainty in the current understanding of MA-tBLG.

\begin{figure}
\includegraphics[clip,width=0.45\textwidth]{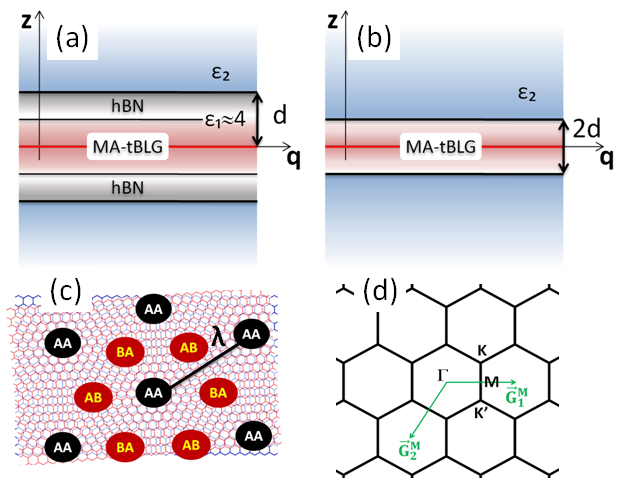}
\vspace{-.30 cm}
\caption{(Color online) Setups for dielectric engineering of MA-tBLG: (a) MA-tBLG encapsulated in hBN, where the dielectric function $\varepsilon_1 \approx 4$ encodes both the contribution from hBN (gray) and from the $\sigma$-band higher energy background of MA-tBLG (red). $\varepsilon_2$ describes the dielectric surrounding (blue) at a distance $d$ from the center of MA-tBLG. MA-tBLG with a metallic gate at a distance $d$ is realized for $\varepsilon_2 \rightarrow \infty$. (b) Dielectric $\varepsilon_2$ in direct contact with MA-tBLG as described by $d=3.35 \: \text{\AA}$, which is the interlayer distance of BLG and hBN \cite{RosPRB922015}. (c) In the moir\'e pattern of MA-tBLG, $AA$ and $AB/BA$ regions can be recognised according to the stacking of the atoms in each graphene layer. The moir\'e lattice constant $\lambda$ corresponds to the distance between adjacent $AA$ regions, and is $134 \: \text{\AA}$ for the first magic angle. (d) The reciprocal space is broken up into the hexagonal mini Brillouin zones (BZ) associated with the moir\'e real-space superlattice. The reciprocal lattice is spanned by the reciprocal lattice vectors $\bold{G}_{1}^M$ and $\bold{G}_{2}^M$ \cite{KosPRX82018}. Neighboring mini-BZ are included in the low-energy continuum description of MA-tBLG up to a certain cutoff $G_c$.}
\label{fig1} 
\vspace{-0.3 cm}
\end{figure}

In this Letter, we provide a quantitative study of the internal polarizability and effects of the dielectric environment on the effective electron-electron interaction in MA-tBLG. We consider MA-tBLG in dielectric surroundings, see Fig. \ref{fig1}(a-b), directly resembling different experimental setups: (a) MA-tBLG is separated by possible hexagonal boron nitride (hBN) encapsulation from a dielectric environment ($\varepsilon_2$) at distance $d$ below and above. The case of a metallic gate at a certain distance $d$ from MA-tBLG is included in this analysis for $\varepsilon_2 \rightarrow \infty$. (b) The special case of the dielectric $\varepsilon_2$ in direct contact with MA-tBLG as realized for $d=3.35 \: \text{\AA}$. Our main findings are: $(1)$ intrinsic screening in MA-tBLG is enhanced by interlayer coupling and is larger than previously assumed in the model of uncoupled graphene double layers \cite{tobias2016,CaoN5562018ins,KosPRX82018,2018arXiv180409047Z,PhysRevLett.121.217001,Guinea13174,PhysRevLett.122.026801,2018arXiv180708190S,2018arXiv180904604F,GonPRB591999}, $(2)$ in a possible Mott insulating phase, the effective interaction can be modulated on the order of $50 \%$ by changing the dielectric environment, and $(3)$ metallic states of MA-tBLG are mostly insensitive to the dielectric environment. Finally, a generically applicable, realistic, and yet simple model of the effective electron-electron interaction in MA-tBLG setups as depicted in Fig. \ref{fig1} is provided.

\textit{Internal screening.} As a first step, we calculate the intrinsic polarization function of MA-tBLG in the random phase approximation (RPA) using the low-energy continuum model for the electronic bands of tBLG from Refs. \cite{BisPNAS1082011,KosPRX82018}. In this model \cite{SM}, two sets of Dirac electron bands originating from the lower and the upper graphene layers hybridize due to the interlayer coupling, which is modulated with the periodicity of the moir\'e superlattice (c.f. Fig. \ref{fig1}(c)). In reciprocal space, there is correspondingly a coupling between Dirac electron states at each $\bold{k}$ vector in one layer with states at $\bold{k}+\bold{G}$ in the other layer, where $\bold{G}$ is a reciprocal lattice vector associated with the moir\'e superlattice: $\bold{G}=m \bold{G}_1^M + n \bold{G}_2^M$ with $m$, $n$ integers and $\bold{G}_1^M$ and $\bold{G}_2^M$ spanning the reciprocal lattice of tBLG, see Fig. \ref{fig1}(d). Then, the electronic states of tBLG are expanded in terms of coupled two-dimensional Dirac spinors up to a certain plane wave cutoff $G_c=8G_M$, where $G_M = |\bold{G}_1^M|=|\bold{G}_2^M|$ \cite{SM}. MA-tBLG is then realized when the twisting angle is set to $1.05 \: ^{\circ}$.

In reciprocal space, the polarization operator $\Pi_0$ is a function of the scattering vector $\bold{q}$ and a matrix indexed by reciprocal lattice vectors $\bold{G}$ and $\bold{G}'$. At zero transferred frequency, $\Pi_0$ takes the form \cite{vonkat_1989}
\begin{equation}
\Pi_0^{\bold{G}, \bold{G}'} (\bold{q}) = \frac{g_s g_v}{S_M N} \sum_{\substack{\bold{k} \\ \alpha , \beta \\ \bold{G}_2, \bold{G}'_2}} \mathcal{M}_{\bold{G}_2, \bold{G}'_2, \bold{G}, \bold{G}'}^{\alpha \beta} \frac{f_\bold{k}^\alpha - f_{\bold{k}+\bold{q}}^{\beta}}{i\eta + E_\bold{k}^\alpha - E_{\bold{k}+\bold{q}}^\beta}.
\label{eq1}
\end{equation}
Here, $g_s=g_v=2$ are the spin and valley degeneracy factors, $S_M=\sqrt{3}\lambda^2/2$ is the moir\'e unit cell area, where $\lambda \approx 134 \: \text{\AA}$ is the moir\'e lattice constant, $N=100$ is the number of $\bold{k}$ points, $\alpha$ and $\beta$ encode the sublattice ($A_1$, $B_1$, $A_2$ and $B_2$) and band indices \cite{SM}, $\bold{G}_2$, $\bold{G}'_2$, $\bold{G}$ and $\bold{G}'$ are reciprocal lattice vectors, $f_\bold{k}^{\alpha}$ and $E_\bold{k}^\alpha$ are, respectively, the Fermi function and the band energy of $\alpha$ at $\bold{k}$, $\eta = 0.5 \cdot 10^{-6}-10^{-8} \: \text{eV}$ is the broadening parameter, and we consider the temperature $T\approx 50$~K (inverse temperature $\beta = 200 \: \text{eV}^{-1}$) to calculate the Fermi functions $f_{k}^\alpha$ in Eq. (\ref{eq1}) \cite{SM}. $\Pi_0$ is thus given in units of $\text{eV}^{-1} \, \text{\AA}^{-2}$. The overlap matrix $\mathcal{M}$ results from the Dirac spinor plane wave expansion coefficients $c_{\bold{k}}^{\alpha, \bold{G}}$ via
 \begin{equation}
\mathcal{M}_{\bold{G}_2, \bold{G}'_2, \bold{G}, \bold{G}'}^{\alpha \beta} = ( c_{\bold{k}}^{\alpha, \bold{G}_2} )^\dagger ( c_{\bold{k}+\bold{q}}^{\beta, \bold{G}'_2-\bold{G}'} )^\dagger c_{\bold{k} + \bold{q}}^{\beta, \bold{G}'_2} c_{\bold{k}}^{\alpha, \bold{G} + \bold{G}_2}.
\label{eq2}
\end{equation}

We then study the internal screening in MA-tBLG for two possible scenarios: first, screening according to the RPA, which assumes MA-tBLG to be in a conventional metallic state; second, a scenario resembling insulating states which are modeled within the so-called constrained RPA (cRPA) \cite{PhysRevB.70.195104}. 
In the cRPA, polarization processes taking place inside the low-energy flat bands are excluded from Eq. (\ref{eq1}) by setting $f_\bold{k}^{\alpha}=\frac{1}{2}$ for all states $\alpha$ from this low-energy sector. The resulting cRPA partially screened interactions are those which should be considered in effective Hamiltonians solely dealing with the low-energy flat bands, and correspond also to the screened interactions that would be expected if electronic correlations suppress low-energy polarization processes like in a Mott insulator \cite{vanLoon_MottScreen,MottScreen}.

The results presented in the following are obtained at charge neutrality with the continuum model from Ref. \cite{KosPRX82018} at a twist angle of $\theta=1.05 \: ^{\circ}$. We show in the supplemental material \cite{SM} that our results are robust against changes in twist angle, doping, and interlayer coupling parameters.



\begin{figure}
\includegraphics[clip,width=0.45\textwidth]{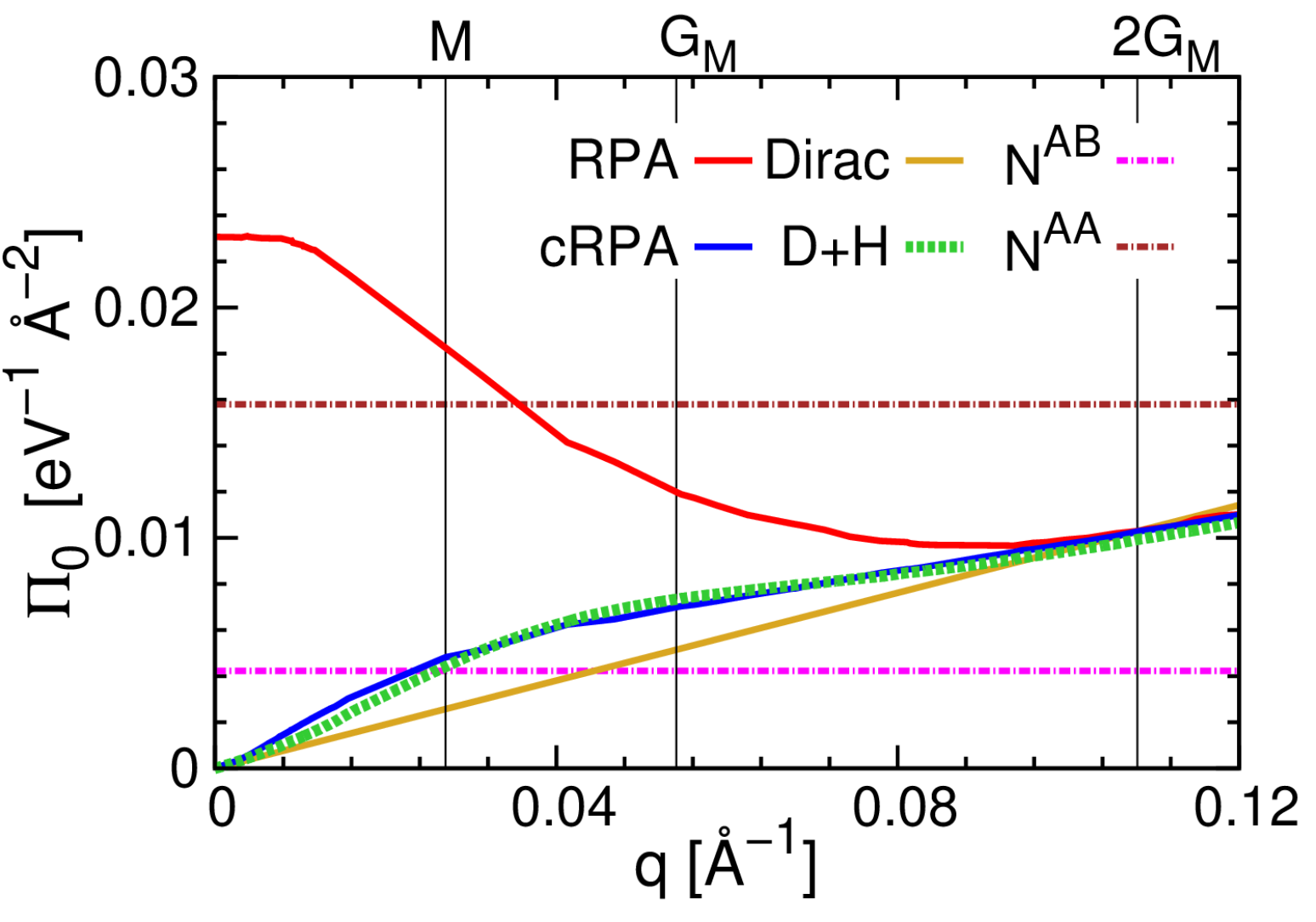}
\vspace{-.30 cm}
\caption{(Color online) Static polarization function of MA-tBLG in comparison to uncoupled tBLG ($\Pi_0^{\rm{Dirac}}$, orange line) from Eq. (\ref{eq:PiDirac}), the model of quasi-localized states plus Dirac background ($\Pi^{\rm{D}+\rm{H}}_0$, green dashed line) from Eq. (\ref{eq:Pi0fit}), and the DOS of commensurate AA (brown dash-dotted line) and AB BLG (pink dash-dotted line). For MA-tBLG, polarizabilities from RPA (red line) assuming a metallic state and cRPA (blue line), i.e. excluding polarization processes within the flat bands similarly to a Mott insulator, are shown. $G_M=0.057 \: \text{\AA}^{-1}$ marks the length of the reciprocal lattice vectors.} 
\label{fig2} 
\vspace{-0.3 cm}
\end{figure}

Fig. \ref{fig2} compares the RPA and cRPA polarization functions \cite{SM,GGpol} for MA-tBLG to the case of uncoupled tBLG, where the interlayer coupling is neglected. Uncoupled tBLG hosts 8 flavors of ideal Dirac fermions due to spin $g_s=2$, valley $g_v=2$, and layer degeneracy $g_l=2$ resulting in a linearly-$q$ dependent polarization function \cite{katsnelson_2012}: 
\begin{equation}
\Pi_0^{\rm Dirac} (q) = g_s g_v g_l \frac{q} { 16 \hbar v_F},
\label{eq:PiDirac}
\end{equation}
where $v_F$ is the Fermi velocity. For $q<G_M$, the intrinsic polarization functions of MA-tBLG, both in RPA and cRPA, systematically exceed the uncoupled tBLG model. Intrinsic screening is, thus, significantly larger in MA-tBLG than previously assumed \cite{tobias2016,CaoN5562018ins,KosPRX82018,2018arXiv180409047Z,PhysRevLett.121.217001,Guinea13174,PhysRevLett.122.026801,2018arXiv180708190S,2018arXiv180904604F,GonPRB591999}. Only at larger momentum transfer $q \gtrsim 1.5 G_M\approx 0.08 \: \text{\AA}^{-1}$, the RPA and cRPA polarization functions of MA-tBLG approach values corresponding to the uncoupled model. This behavior can be understood from the interplay of intra- and interlayer coupling: at sufficiently large momentum transfer $q > G_M$, the intralayer coupling $\hbar v_F |q| \gtrsim 0.3 \: \text{eV}$ dominates over the interlayer one. At $q<G_M$, however, the interlayer coupling cannot be neglected.

In the RPA model, i.e. in the metallic case, the enhancement of screening for $q<G_M$ is due to the high density of states (DOS) around the Fermi level originating from the flat bands. In the cRPA screening processes inside flat bands are excluded, which explains why the polarization function is smaller than in the RPA case. Still, the cRPA polarizability is clearly enhanced as compared to the uncoupled case for $q<G_M$. This can be understood as follows: While polarization processes taking place entirely inside the flat band manifold are excluded in cRPA, gapped transitions between states associated with peaks in the DOS, e.g. between the flat bands and higher energy states with corresponding peaks in the MA-tBLG DOS around $E\approx\pm 20$\,meV and $\pm 60\,$meV \cite{SM} are possible. These gapped transitions are reminiscent of atomic systems. The hydrogen atom, as simplest example, would yield a q-dependent polarizability of the form $\Pi_0^{\rm{H}} (q)\sim \frac{q^2}{\left( 1 + \left( \frac{q}{b} \right)^2 \right)^5}$ \cite{PhysRevA.43.1186}, where the parameter $b$ is an effective inverse orbital radius. Superimposing this quasi-atomic model $\Pi_0^{\rm{H}}$ with a Dirac electron background capturing all higher energy processes leads to the ansatz: 
\begin{equation}
\Pi^{\rm{D}+\rm{H}}_0 (q) = \Pi^{\rm{Dirac}}_0 (q) + a \frac{q^2}{\left( 1 + \left( \frac{q}{b} \right)^2 \right)^5}.
\label{eq:Pi0fit}
\end{equation}
Fig. \ref{fig2} shows that Eq. \ref{eq:Pi0fit} fits the cRPA numerics very well, upon choosing $a\approx 4.1 \: \text{eV}^{-1}$ and $b \approx 0.088 \: \text{\AA}^{-1}$. Thus, we suggest to use Eq. (\ref{eq:Pi0fit}) for calculations of cRPA screened Coulomb interaction matrix elements in MA-tBLG.

While screening in MA-tBLG has been often approximated in terms of uncoupled tBLG \cite{CaoN5562018ins,KosPRX82018,2018arXiv180409047Z,PhysRevLett.121.217001,Guinea13174,PhysRevLett.122.026801,2018arXiv180708190S,2018arXiv180904604F,GonPRB591999}, an alternative point of view is that MA-tBLG is a patchwork of AA and AB stacked bilayer graphene (BLG) regions. For $q\gtrsim G_M/2$ the MA-tBLG polarizability fits, indeed, into the range marked by the low-energy density of states (DOS) of AA BLG, $N^{AA}(E_F) = 0.0158 \: \text{eV}^{-1} \, \text{\AA}^{-2}$, and AB BLG, $N^{AB}(E_F) = 0.0042 \: \text{eV}^{-1} \, \text{\AA}^{-2}$ \cite{HwaPRL1012008}. Further below we show that the AB BLG model gives good estimates for the effective local interactions when compared with the cRPA. On physical grounds, our results for the polarization function show that MA-tBLG can be seen as a flat band system embedded in an almost metallic background generated by the higher energy bands, which are separated by approximately $15 \: \text{meV}$ from the low-energy flat bands \cite{SM}.

\textit{Dielectric engineering.} We now assess the possibilities for dielectric engineering of MA-tBLG in experimental setups as shown in Fig. \ref{fig1}(a-b). The screened interaction is
\begin{equation}
W(q)= \frac{V(q)}{\varepsilon (q)},
\label{eq:scint}
\end{equation}
where $V(q)=2\pi e^2/q$ is the bare interaction and $\varepsilon (q)= \varepsilon_{\text{env}} (q) + V(q) \Pi_0 (q)$ is the dielectric function \cite{katsnelson_2012}. The latter accounts for the screening resulting from the electrons in the bands of MA-tBLG via $\Pi_0 (q)$ plus the screening produced by the dielectric surrounding and higher energy bands of MA-tBLG summarized in $\varepsilon_{\text{env}} (q)$. In the setups from Fig. \ref{fig1}, the background dielectric function reads \cite{RosPRB922015}
\begin{equation}
\varepsilon_{\text{env}} (q) = \varepsilon_1 \frac{ 1 - \tilde{\varepsilon}_2^2 e^{-4 q d}}{\left( 1 + \tilde{\varepsilon}_2 e^{-2 q d} \right)^2},
\label{eq:eps_eff}
\end{equation}
where we approximated $\varepsilon_1 \approx 4$ \cite{katsnelson_2012} and $\tilde{\varepsilon}_2 = (\varepsilon_1-\varepsilon_2)/(\varepsilon_1+\varepsilon_2)$. Together with the polarizability $\Pi_0(q)$ obtained either numerically in cRPA from Eq. (\ref{eq1}) or conveniently from the fit of Eq. (\ref{eq:Pi0fit}), Eqs. (\ref{eq:scint}) and (\ref{eq:eps_eff}) specify the appropriately screened interaction of MA-tBLG in different dielectric surroundings, which should be incorporated in interacting electron models of the low-energy flat bands.

To illustrate the influence of the dielectric surrounding on the effective interaction in MA-tBLG in terms of a single descriptor, we consider an effective local interaction $U$ obtained by Fourier transformation of $W(q)$: 
\begin{equation}
U=\int \frac{d^2 \bold{q}}{(2 \pi)^2} W(\bold{q}) = \int_{0}^{q_c} \frac{dq}{2 \pi} q W(q).
\label{eq3}
\end{equation}
Here, the cutoff $q_c$ relates to the 2D radial spread $r_{WF}$ of the Wannier functions (WF) constructed from the MA-tBLG low-energy bands: $q_c \approx \pi/r_{WF}$. We choose $r_{WF}$ to be half of the moir\'e lattice constant $\lambda$, hence $q_c \approx 0.047 \: \text{\AA}^{-1}$.


\begin{figure*}
\leavevmode
\includegraphics[clip,width=0.325\textwidth]{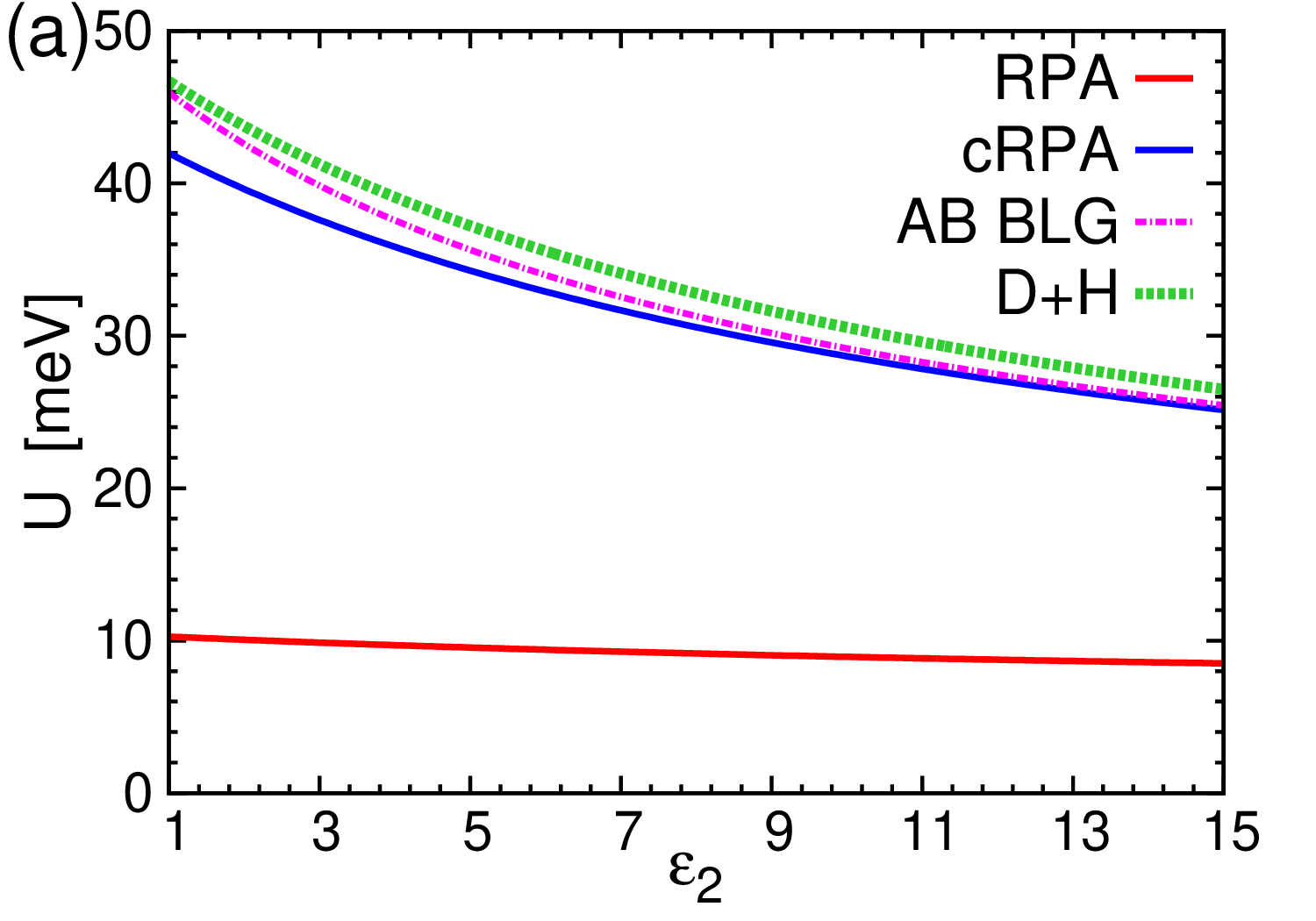}
\includegraphics[clip,width=0.325\textwidth]{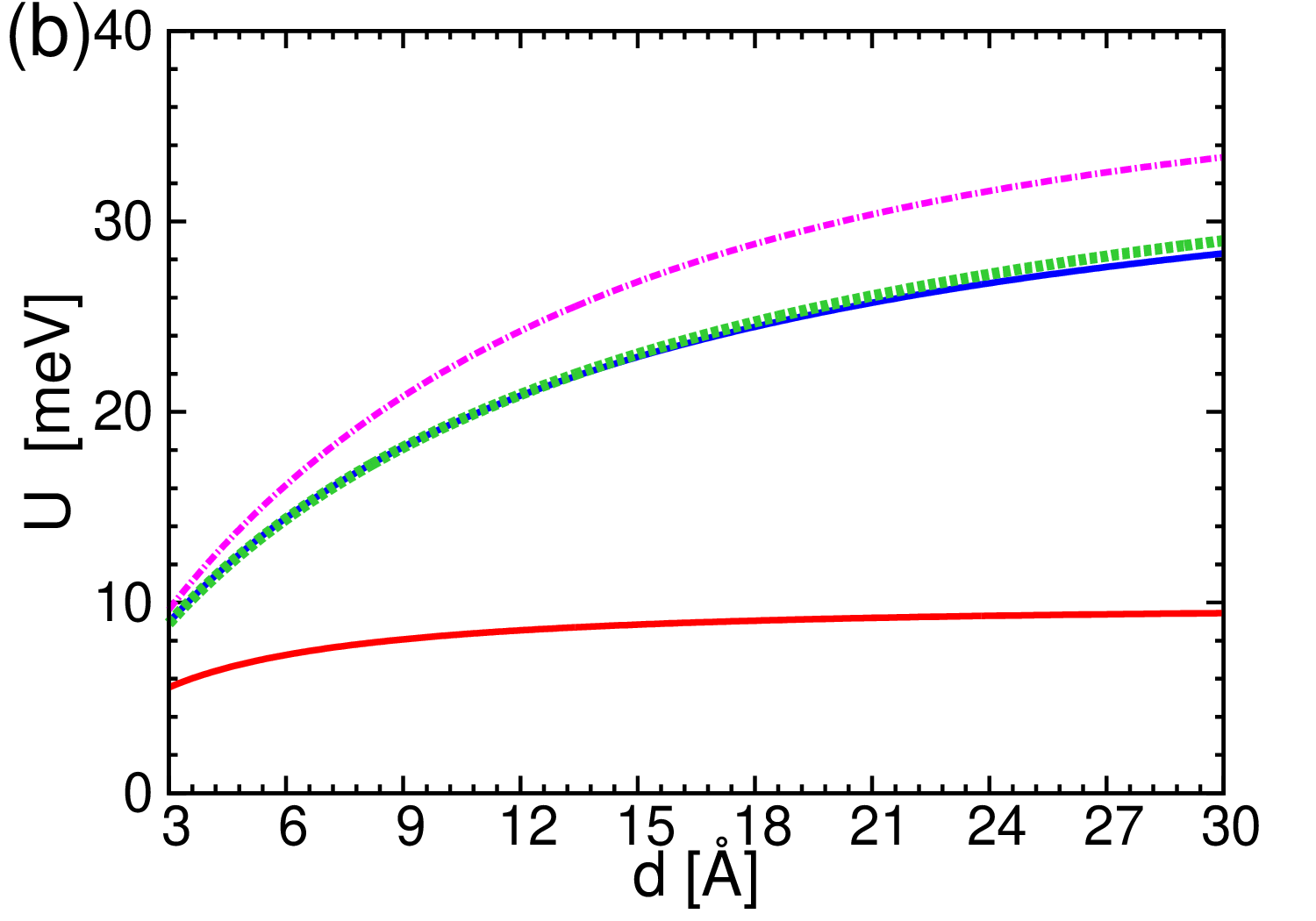} 
\includegraphics[clip,width=0.325\textwidth]{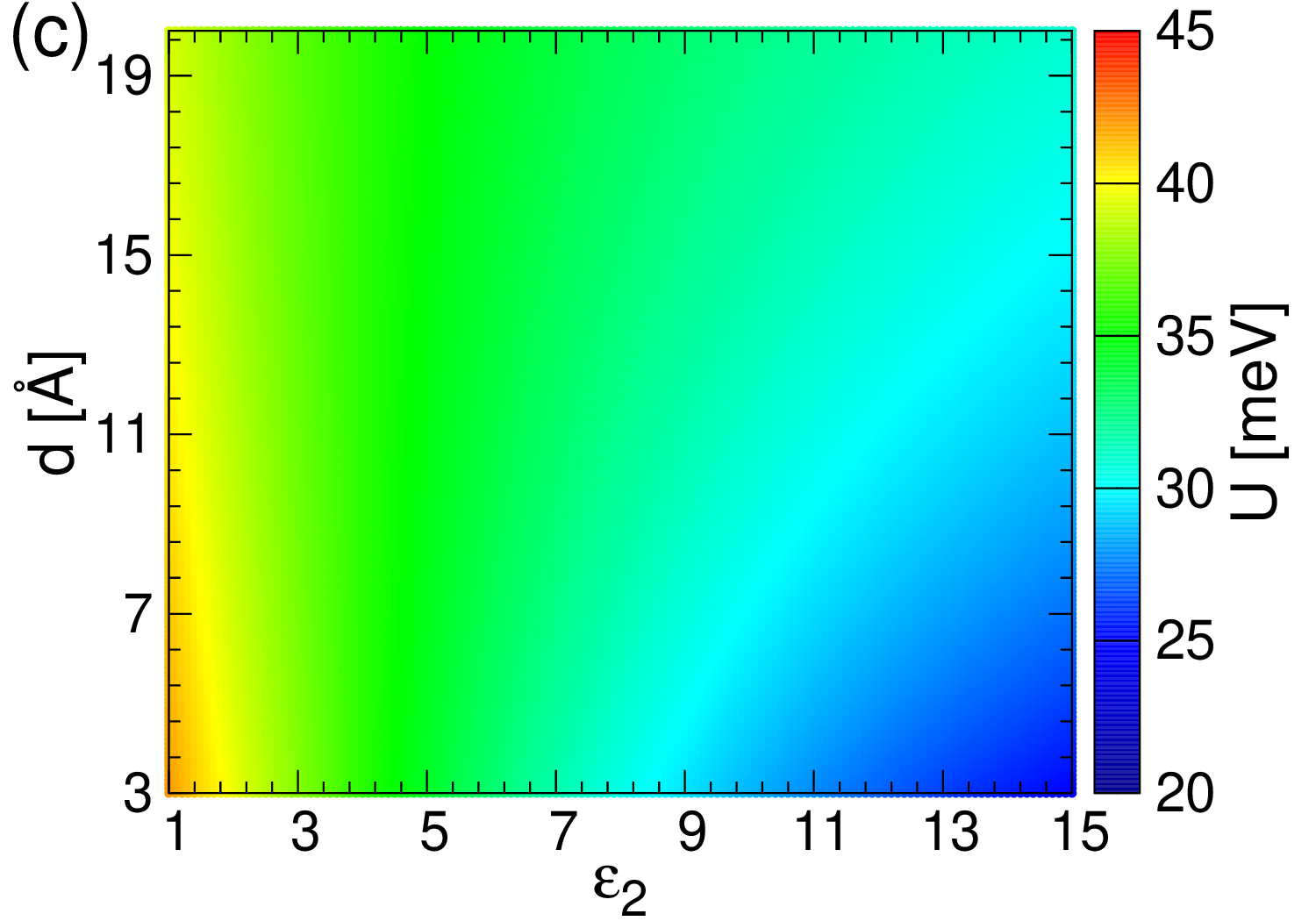} 
\caption{(Color online) Dielectric engineering of MA-tBLG with an external dielectric $\varepsilon_2$ at a distance $d$. (a) Effective local interaction $U$ as a function of the dielectric constant $\varepsilon_2$ of the environment in direct contact with MA-tBLG. The RPA (red line) mimicking metallic screening and cRPA (blue line) numerical results assuming suppressed low energy screening for MA-tBLG are compared with the analytical estimates from the $AB$ BLG model (pink dash-dotted line) and the quasi-atomic model with Dirac background (D+H, green dashed line) from Eq. (\ref{eq:Pi0fit}). (b) Influence of a metallic gate ($\varepsilon_2 \rightarrow \infty$) at a distance $d$ from MA-tBLG on $U$. (c) Effective cRPA screened local interaction $U$ (color coded) as a function of the dielectric surrounding $\varepsilon_2$ and its distance $d$ from MA-tBLG.} 
\label{fig3} 
\vspace{-0.3cm}
\end{figure*}

Fig. \ref{fig3}(a) shows the influence of the dielectric environment $\varepsilon_2$ on $U$, where we assume that the surrounding dielectric is in direct contact with MA-tBLG (c.f. Fig. \ref{fig1}(b)). Assuming a metallic state of MA-tBLG as in the RPA model, the dielectric environment has barely any effects, and $U$ remains almost constant around $10 \: \text{meV}$. This insensitivity of metallic states in MA-tBLG to the dielectric surrounding is due to a high intrinsic polarizability which masks any environmental polarizabilities (c.f. RPA curves in Fig. \ref{fig2} and Fig. S3 of the supplemental material \cite{SM}).

On the other hand, the cRPA estimated $U$ varies from $40 \: \text{meV}$ to $\approx 20 \: \text{meV}$ when $\varepsilon_2$ increases. Interestingly, $40 \: \text{meV}$ is approximately the splitting of the upper and lower Hubbard bands when a gap is opened at the CNP, as recently measured in scanning tunneling spectroscopy (STS) \cite{2019arXiv190102997C}. 
Besides its relevance for low-energy effective Hamiltonians, the cRPA solution resembles the fully screened interaction in certain insulating states, where screening from the low-energy states is suppressed. In this situation the dielectric environment can modify the effective interaction $U$ up to $\approx 40-50 \, \%$.

The influence of the metallic gate ($\varepsilon_2 \rightarrow \infty$) at a distance $d$ from MA-tBLG (c.f. Fig. \ref{fig1}(b)) on the effective local interaction $U$ is shown in Fig. \ref{fig3}(b). In an insulating state (cRPA case), the influence of the metallic gate is strong if the distance from MA-tBLG is on the order of $d \lesssim 20 \: \text{\AA}$. In this case, $U$ is reduced by a factor of two when approaching the gate from $d \sim 30 \: \text{\AA}$ to $d \sim 10 \: \text{\AA}$. The fact that the metallic gate has to be as close as $d \lesssim 20 \: \text{\AA}$, which is much smaller than the moir\'e lattice constant ($134 \: \text{\AA}$), in order to see any substancial effect on $U$ even in the cRPA / insulating case is due to MA-tBLG being a flat-band system embedded in an already almost metallic background. In the RPA case, the metallic gate has again a much smaller effect on $U$. Only for metallic gates very close to MA-tBLG ($d \approx 3.35 \: \text{\AA}$), $U$ is reduced from $10 \: \text{meV}$ to $5 \: \text{meV}$. Thus, in metallic states of MA-tBLG (RPA case), metallic gates can only influence the electron-electron interactions due to external screening if they are very close to the MA-tBLG, and a possible hBN spacer separating the gate from the MA-tBLG should be as thin as a monolayer. The dependence of the effective local cRPA screened interaction $U$ on the dielectric surrounding $\varepsilon_2$ and its distance $d$ from MA-tBLG is summarized in Fig. \ref{fig3}(c). $U$ can be reduced from $40 \: \text{meV}$ (free standing case) to $10 \: \text{meV}$ (strong screening environment) if the surrounding dielectric gets sufficiently close $d \lesssim 20 \: \text{\AA}$.

Additionally to the RPA and cRPA numerics according to Eq. (\ref{eq1}), we also propose simple approximate forms to calculate screened Coulomb interaction matrix elements. First, by choosing the expression of Eq. (\ref{eq:Pi0fit}) for the polarizability, we reproduce also the effective cRPA screened local interaction quite accurately. Even more simplistically, Fig. \ref{fig3} also shows that screening based on the DOS of $AB$ BLG provides us with a good estimate for $U$ in the cRPA case. 

\textit{Summary.} We have shown that the sensitivity of the interaction profile in MA-tBLG to the dielectric environment depends decisively on the quantum state of the electrons in the low-energy flat-band manifold.  Currently,  the nature of the insulating states at half-filling, of superconductivity near half-filling, and of the gaps measured in STS at charge neutrality are major open problems in the understanding of MA-tBLG. 
What would an experimentally detected significant influence of the dielectric surrounding on spectral, thermodynamic, or transport properties in any of these phases imply? Based on our results any strong influence of the external dielectric means that charge screening in the low-energy bands is substantially suppressed as compared to a conventional metallic state as assumed in RPA. That is hardly possible for insulating or any kind of ordered states resulting from a weak coupling instability, which affect only a small sector of states in the vicinity of the non-interacting Fermi surface. Thus, significant impact of external dielectrics on correlated insulating states implies that they are at strong coupling in the sense that states from the entire mini-BZ contribute significantly. Similarly, any impact of the dielectric surrounding on superconductivity implies strong coupling physics at work, and possibly hints at an unconventional origin.

Currently, there are arguments in favor and against strong coupling physics governing different parts of the MA-tBLG phase diagram. On the pro side, is the ratio of effective interactions and bandwidth: in case of partial screening as described in cRPA, the effective interaction largely exceeds the bandwidth, which is suggestive of strong coupling. In line with a strong coupling scenario, close to charge neutrality, STS further measured "gaps" on the scale of several $10\,$meV \cite{2019arXiv190102997C} and transport experiments reported suppression of conductance around charge neutrality being robust against temperature and magnetic fields \cite{Yankowitz1059,CaoN5562018ins,2019arXiv190306513L}. At and around half-filling, the situation is more complex: the generic shape of the phase diagram in this region appears to exhibit similarities to the cuprate high-$T_c$ superconductors, which might be seen as an indication pro strong coupling physics. At half filling, however, the insulating states are fragile, i.e. they are easily destroyed by magnetic fields ($B \sim 5 \: \text{T}$) and temperature ($T \sim 4 \: \text{K}$), in line with gaps that are small ($\Delta \sim 0.3 \: \text{meV}$) \cite{Yankowitz1059,CaoN5562018ins,2019arXiv190306513L}. All these energies are at least an order of magnitude smaller than the expected width of the flat bands, and would be consistent with a weak coupling scenario. Experimental studies of the response of MA-tBLG to changes in the surrounding dielectric will thus be useful to distinguish between strong- and weak-coupling scenarios, and to pinpoint the nature of the different quantum many-body states in MA-tBLG. 

We thank D. A. Abanin, E. Bascones, F. Guinea for useful conversations. JMP and TW acknowledge funding from DFG-RTG 2247 (QM$^3$) and the European Graphene Flagship. RT is supported by DFG-SFB 1170 Tocotronics (project B04), DFG-SPP 1666, and further acknowledges financial support from the DFG through the W\"urzburg-Dresden Cluster of Excellence on Complexity and Topology in Quantum Matter -- \textit{ct.qmat} (EXC 2147, project-id 39085490). RV acknowledges funding from DFG-TRR 49.

\bibliography{tblg_bib}

\begin{thebibliography}{52}
\expandafter\ifx\csname natexlab\endcsname\relax\def\natexlab#1{#1}\fi
\expandafter\ifx\csname bibnamefont\endcsname\relax
  \def\bibnamefont#1{#1}\fi
\expandafter\ifx\csname bibfnamefont\endcsname\relax
  \def\bibfnamefont#1{#1}\fi
\expandafter\ifx\csname citenamefont\endcsname\relax
  \def\citenamefont#1{#1}\fi
\expandafter\ifx\csname url\endcsname\relax
  \def\url#1{\texttt{#1}}\fi
\expandafter\ifx\csname urlprefix\endcsname\relax\def\urlprefix{URL }\fi
\providecommand{\bibinfo}[2]{#2}
\providecommand{\eprint}[2][]{\url{#2}}

\bibitem[{\citenamefont{Su\'arez~Morell
  et~al.}(2010)\citenamefont{Su\'arez~Morell, Correa, Vargas, Pacheco, and
  Barticevic}}]{SuaPRB822010}
\bibinfo{author}{\bibfnamefont{E.}~\bibnamefont{Su\'arez~Morell}},
  \bibinfo{author}{\bibfnamefont{J.~D.} \bibnamefont{Correa}},
  \bibinfo{author}{\bibfnamefont{P.}~\bibnamefont{Vargas}},
  \bibinfo{author}{\bibfnamefont{M.}~\bibnamefont{Pacheco}}, \bibnamefont{and}
  \bibinfo{author}{\bibfnamefont{Z.}~\bibnamefont{Barticevic}},
  \bibinfo{journal}{Phys. Rev. B} \textbf{\bibinfo{volume}{82}},
  \bibinfo{pages}{121407(R)} (\bibinfo{year}{2010}).

\bibitem[{\citenamefont{Bistritzer and MacDonald}(2011)}]{BisPNAS1082011}
\bibinfo{author}{\bibfnamefont{R.}~\bibnamefont{Bistritzer}} \bibnamefont{and}
  \bibinfo{author}{\bibfnamefont{A.~H.} \bibnamefont{MacDonald}},
  \bibinfo{journal}{Proceedings of the National Academy of Sciences}
  \textbf{\bibinfo{volume}{108}}, \bibinfo{pages}{12233}
  (\bibinfo{year}{2011}).

\bibitem[{\citenamefont{Cao et~al.}(2018{\natexlab{a}})\citenamefont{Cao,
  Fatemi, Demir, Fang, Tomarken, Luo, Sanchez-Yamagishi, Watanabe, Taniguchi,
  Kaxiras et~al.}}]{CaoN5562018ins}
\bibinfo{author}{\bibfnamefont{Y.}~\bibnamefont{Cao}},
  \bibinfo{author}{\bibfnamefont{V.}~\bibnamefont{Fatemi}},
  \bibinfo{author}{\bibfnamefont{A.}~\bibnamefont{Demir}},
  \bibinfo{author}{\bibfnamefont{S.}~\bibnamefont{Fang}},
  \bibinfo{author}{\bibfnamefont{S.~L.} \bibnamefont{Tomarken}},
  \bibinfo{author}{\bibfnamefont{J.~Y.} \bibnamefont{Luo}},
  \bibinfo{author}{\bibfnamefont{J.~D.} \bibnamefont{Sanchez-Yamagishi}},
  \bibinfo{author}{\bibfnamefont{K.}~\bibnamefont{Watanabe}},
  \bibinfo{author}{\bibfnamefont{T.}~\bibnamefont{Taniguchi}},
  \bibinfo{author}{\bibfnamefont{E.}~\bibnamefont{Kaxiras}},
  \bibnamefont{et~al.}, \bibinfo{journal}{Nature}
  \textbf{\bibinfo{volume}{556}}, \bibinfo{pages}{80}
  (\bibinfo{year}{2018}{\natexlab{a}}).

\bibitem[{\citenamefont{Cao et~al.}(2018{\natexlab{b}})\citenamefont{Cao,
  Fatemi, Fang, Watanabe, Taniguchi, Kaxiras, and
  Jarillo-Herrero}}]{CaoN5562018sc}
\bibinfo{author}{\bibfnamefont{Y.}~\bibnamefont{Cao}},
  \bibinfo{author}{\bibfnamefont{V.}~\bibnamefont{Fatemi}},
  \bibinfo{author}{\bibfnamefont{S.}~\bibnamefont{Fang}},
  \bibinfo{author}{\bibfnamefont{K.}~\bibnamefont{Watanabe}},
  \bibinfo{author}{\bibfnamefont{T.}~\bibnamefont{Taniguchi}},
  \bibinfo{author}{\bibfnamefont{E.}~\bibnamefont{Kaxiras}}, \bibnamefont{and}
  \bibinfo{author}{\bibfnamefont{P.}~\bibnamefont{Jarillo-Herrero}},
  \bibinfo{journal}{Nature} \textbf{\bibinfo{volume}{556}}, \bibinfo{pages}{43}
  (\bibinfo{year}{2018}{\natexlab{b}}).

\bibitem[{\citenamefont{Yankowitz et~al.}(2019)\citenamefont{Yankowitz, Chen,
  Polshyn, Zhang, Watanabe, Taniguchi, Graf, Young, and Dean}}]{Yankowitz1059}
\bibinfo{author}{\bibfnamefont{M.}~\bibnamefont{Yankowitz}},
  \bibinfo{author}{\bibfnamefont{S.}~\bibnamefont{Chen}},
  \bibinfo{author}{\bibfnamefont{H.}~\bibnamefont{Polshyn}},
  \bibinfo{author}{\bibfnamefont{Y.}~\bibnamefont{Zhang}},
  \bibinfo{author}{\bibfnamefont{K.}~\bibnamefont{Watanabe}},
  \bibinfo{author}{\bibfnamefont{T.}~\bibnamefont{Taniguchi}},
  \bibinfo{author}{\bibfnamefont{D.}~\bibnamefont{Graf}},
  \bibinfo{author}{\bibfnamefont{A.~F.} \bibnamefont{Young}}, \bibnamefont{and}
  \bibinfo{author}{\bibfnamefont{C.~R.} \bibnamefont{Dean}},
  \bibinfo{journal}{Science} \textbf{\bibinfo{volume}{363}},
  \bibinfo{pages}{1059} (\bibinfo{year}{2019}).

\bibitem[{\citenamefont{{Choi} et~al.}(2019)\citenamefont{{Choi}, {Kemmer},
  {Peng}, {Thomson}, {Arora}, {Polski}, {Zhang}, {Ren}, {Alicea}, {Refael}
  et~al.}}]{2019arXiv190102997C}
\bibinfo{author}{\bibfnamefont{Y.}~\bibnamefont{{Choi}}},
  \bibinfo{author}{\bibfnamefont{J.}~\bibnamefont{{Kemmer}}},
  \bibinfo{author}{\bibfnamefont{Y.}~\bibnamefont{{Peng}}},
  \bibinfo{author}{\bibfnamefont{A.}~\bibnamefont{{Thomson}}},
  \bibinfo{author}{\bibfnamefont{H.}~\bibnamefont{{Arora}}},
  \bibinfo{author}{\bibfnamefont{R.}~\bibnamefont{{Polski}}},
  \bibinfo{author}{\bibfnamefont{Y.}~\bibnamefont{{Zhang}}},
  \bibinfo{author}{\bibfnamefont{H.}~\bibnamefont{{Ren}}},
  \bibinfo{author}{\bibfnamefont{J.}~\bibnamefont{{Alicea}}},
  \bibinfo{author}{\bibfnamefont{G.}~\bibnamefont{{Refael}}},
  \bibnamefont{et~al.}, \bibinfo{journal}{arXiv 1901.02997}
  (\bibinfo{year}{2019}).

\bibitem[{\citenamefont{{Lu} et~al.}(2019)\citenamefont{{Lu}, {Stepanov},
  {Yang}, {Xie}, {Aamir}, {Das}, {Urgell}, {Watanabe}, {Taniguchi}, {Zhang}
  et~al.}}]{2019arXiv190306513L}
\bibinfo{author}{\bibfnamefont{X.}~\bibnamefont{{Lu}}},
  \bibinfo{author}{\bibfnamefont{P.}~\bibnamefont{{Stepanov}}},
  \bibinfo{author}{\bibfnamefont{W.}~\bibnamefont{{Yang}}},
  \bibinfo{author}{\bibfnamefont{M.}~\bibnamefont{{Xie}}},
  \bibinfo{author}{\bibfnamefont{M.~A.} \bibnamefont{{Aamir}}},
  \bibinfo{author}{\bibfnamefont{I.}~\bibnamefont{{Das}}},
  \bibinfo{author}{\bibfnamefont{C.}~\bibnamefont{{Urgell}}},
  \bibinfo{author}{\bibfnamefont{K.}~\bibnamefont{{Watanabe}}},
  \bibinfo{author}{\bibfnamefont{T.}~\bibnamefont{{Taniguchi}}},
  \bibinfo{author}{\bibfnamefont{G.}~\bibnamefont{{Zhang}}},
  \bibnamefont{et~al.}, \bibinfo{journal}{arXiv e-prints}
  (\bibinfo{year}{2019}), \eprint{1903.06513}.

\bibitem[{\citenamefont{Koshino et~al.}(2018)\citenamefont{Koshino, Yuan,
  Koretsune, Ochi, Kuroki, and Fu}}]{KosPRX82018}
\bibinfo{author}{\bibfnamefont{M.}~\bibnamefont{Koshino}},
  \bibinfo{author}{\bibfnamefont{N.~F.~Q.} \bibnamefont{Yuan}},
  \bibinfo{author}{\bibfnamefont{T.}~\bibnamefont{Koretsune}},
  \bibinfo{author}{\bibfnamefont{M.}~\bibnamefont{Ochi}},
  \bibinfo{author}{\bibfnamefont{K.}~\bibnamefont{Kuroki}}, \bibnamefont{and}
  \bibinfo{author}{\bibfnamefont{L.}~\bibnamefont{Fu}}, \bibinfo{journal}{Phys.
  Rev. X} \textbf{\bibinfo{volume}{8}}, \bibinfo{pages}{031087}
  (\bibinfo{year}{2018}).

\bibitem[{\citenamefont{Rademaker and Mellado}(2018)}]{RadPRB982018}
\bibinfo{author}{\bibfnamefont{L.}~\bibnamefont{Rademaker}} \bibnamefont{and}
  \bibinfo{author}{\bibfnamefont{P.}~\bibnamefont{Mellado}},
  \bibinfo{journal}{Phys. Rev. B} \textbf{\bibinfo{volume}{98}},
  \bibinfo{pages}{235158} (\bibinfo{year}{2018}).

\bibitem[{\citenamefont{Pizarro et~al.}(2019)\citenamefont{Pizarro,
  Calder{\'{o}}n, and Bascones}}]{Pizarro_2019}
\bibinfo{author}{\bibfnamefont{J.~M.} \bibnamefont{Pizarro}},
  \bibinfo{author}{\bibfnamefont{M.~J.} \bibnamefont{Calder{\'{o}}n}},
  \bibnamefont{and} \bibinfo{author}{\bibfnamefont{E.}~\bibnamefont{Bascones}},
  \bibinfo{journal}{Journal of Physics Communications}
  \textbf{\bibinfo{volume}{3}}, \bibinfo{pages}{035024} (\bibinfo{year}{2019}).

\bibitem[{\citenamefont{Keimer et~al.}(2015)\citenamefont{Keimer, Kivelson,
  Norman, Uchida, and Zaanen}}]{KeiN5182015}
\bibinfo{author}{\bibfnamefont{B.}~\bibnamefont{Keimer}},
  \bibinfo{author}{\bibfnamefont{S.~A.} \bibnamefont{Kivelson}},
  \bibinfo{author}{\bibfnamefont{M.~R.} \bibnamefont{Norman}},
  \bibinfo{author}{\bibfnamefont{S.}~\bibnamefont{Uchida}}, \bibnamefont{and}
  \bibinfo{author}{\bibfnamefont{J.}~\bibnamefont{Zaanen}},
  \bibinfo{journal}{Nature} \textbf{\bibinfo{volume}{518}}
  (\bibinfo{year}{2015}).

\bibitem[{\citenamefont{Kordyuk}(2015)}]{KorLTP412015}
\bibinfo{author}{\bibfnamefont{A.}~\bibnamefont{Kordyuk}},
  \bibinfo{journal}{Low Temperature Physics} \textbf{\bibinfo{volume}{41}},
  \bibinfo{pages}{319} (\bibinfo{year}{2015}).

\bibitem[{\citenamefont{Kurosaki et~al.}(2005)\citenamefont{Kurosaki, Shimizu,
  Miyagawa, Kanoda, and Saito}}]{KurPRL952005}
\bibinfo{author}{\bibfnamefont{Y.}~\bibnamefont{Kurosaki}},
  \bibinfo{author}{\bibfnamefont{Y.}~\bibnamefont{Shimizu}},
  \bibinfo{author}{\bibfnamefont{K.}~\bibnamefont{Miyagawa}},
  \bibinfo{author}{\bibfnamefont{K.}~\bibnamefont{Kanoda}}, \bibnamefont{and}
  \bibinfo{author}{\bibfnamefont{G.}~\bibnamefont{Saito}},
  \bibinfo{journal}{Phys. Rev. Lett.} \textbf{\bibinfo{volume}{95}},
  \bibinfo{pages}{177001} (\bibinfo{year}{2005}).

\bibitem[{\citenamefont{Sipos et~al.}(2008)\citenamefont{Sipos, Kusmartseva,
  Akrap, Berger, Forr{\'o}, and Tutis}}]{SipNM72008}
\bibinfo{author}{\bibfnamefont{B.}~\bibnamefont{Sipos}},
  \bibinfo{author}{\bibfnamefont{A.~F.} \bibnamefont{Kusmartseva}},
  \bibinfo{author}{\bibfnamefont{A.}~\bibnamefont{Akrap}},
  \bibinfo{author}{\bibfnamefont{H.}~\bibnamefont{Berger}},
  \bibinfo{author}{\bibfnamefont{L.}~\bibnamefont{Forr{\'o}}},
  \bibnamefont{and} \bibinfo{author}{\bibfnamefont{E.}~\bibnamefont{Tutis}},
  \bibinfo{journal}{Nature Materials} \textbf{\bibinfo{volume}{7}}
  (\bibinfo{year}{2008}).

\bibitem[{\citenamefont{Po et~al.}(2018)\citenamefont{Po, Zou, Vishwanath, and
  Senthil}}]{PhysRevX.8.031089}
\bibinfo{author}{\bibfnamefont{H.~C.} \bibnamefont{Po}},
  \bibinfo{author}{\bibfnamefont{L.}~\bibnamefont{Zou}},
  \bibinfo{author}{\bibfnamefont{A.}~\bibnamefont{Vishwanath}},
  \bibnamefont{and} \bibinfo{author}{\bibfnamefont{T.}~\bibnamefont{Senthil}},
  \bibinfo{journal}{Phys. Rev. X} \textbf{\bibinfo{volume}{8}},
  \bibinfo{pages}{031089} (\bibinfo{year}{2018}).

\bibitem[{\citenamefont{Roy and Juri\ifmmode \check{c}\else
  \v{c}\fi{}i\ifmmode~\acute{c}\else \'{c}\fi{}}(2019)}]{2018arXiv180311190R}
\bibinfo{author}{\bibfnamefont{B.}~\bibnamefont{Roy}} \bibnamefont{and}
  \bibinfo{author}{\bibfnamefont{V.}~\bibnamefont{Juri\ifmmode \check{c}\else
  \v{c}\fi{}i\ifmmode~\acute{c}\else \'{c}\fi{}}}, \bibinfo{journal}{Phys. Rev.
  B} \textbf{\bibinfo{volume}{99}}, \bibinfo{pages}{121407(R)}
  (\bibinfo{year}{2019}).

\bibitem[{\citenamefont{Dodaro et~al.}(2018)\citenamefont{Dodaro, Kivelson,
  Schattner, Sun, and Wang}}]{PhysRevB.98.075154}
\bibinfo{author}{\bibfnamefont{J.~F.} \bibnamefont{Dodaro}},
  \bibinfo{author}{\bibfnamefont{S.~A.} \bibnamefont{Kivelson}},
  \bibinfo{author}{\bibfnamefont{Y.}~\bibnamefont{Schattner}},
  \bibinfo{author}{\bibfnamefont{X.~Q.} \bibnamefont{Sun}}, \bibnamefont{and}
  \bibinfo{author}{\bibfnamefont{C.}~\bibnamefont{Wang}},
  \bibinfo{journal}{Phys. Rev. B} \textbf{\bibinfo{volume}{98}},
  \bibinfo{pages}{075154} (\bibinfo{year}{2018}).

\bibitem[{\citenamefont{Liu et~al.}(2018)\citenamefont{Liu, Zhang, Chen, and
  Yang}}]{PhysRevLett.121.217001}
\bibinfo{author}{\bibfnamefont{C.-C.} \bibnamefont{Liu}},
  \bibinfo{author}{\bibfnamefont{L.-D.} \bibnamefont{Zhang}},
  \bibinfo{author}{\bibfnamefont{W.-Q.} \bibnamefont{Chen}}, \bibnamefont{and}
  \bibinfo{author}{\bibfnamefont{F.}~\bibnamefont{Yang}},
  \bibinfo{journal}{Phys. Rev. Lett.} \textbf{\bibinfo{volume}{121}},
  \bibinfo{pages}{217001} (\bibinfo{year}{2018}).

\bibitem[{\citenamefont{Kennes et~al.}(2018)\citenamefont{Kennes, Lischner, and
  Karrasch}}]{PhysRevB.98.241407}
\bibinfo{author}{\bibfnamefont{D.~M.} \bibnamefont{Kennes}},
  \bibinfo{author}{\bibfnamefont{J.}~\bibnamefont{Lischner}}, \bibnamefont{and}
  \bibinfo{author}{\bibfnamefont{C.}~\bibnamefont{Karrasch}},
  \bibinfo{journal}{Phys. Rev. B} \textbf{\bibinfo{volume}{98}},
  \bibinfo{pages}{241407(R)} (\bibinfo{year}{2018}).

\bibitem[{\citenamefont{Isobe et~al.}(2018)\citenamefont{Isobe, Yuan, and
  Fu}}]{PhysRevX.8.041041}
\bibinfo{author}{\bibfnamefont{H.}~\bibnamefont{Isobe}},
  \bibinfo{author}{\bibfnamefont{N.~F.~Q.} \bibnamefont{Yuan}},
  \bibnamefont{and} \bibinfo{author}{\bibfnamefont{L.}~\bibnamefont{Fu}},
  \bibinfo{journal}{Phys. Rev. X} \textbf{\bibinfo{volume}{8}},
  \bibinfo{pages}{041041} (\bibinfo{year}{2018}).

\bibitem[{\citenamefont{Wu et~al.}(2018)\citenamefont{Wu, MacDonald, and
  Martin}}]{PhysRevLett.121.257001}
\bibinfo{author}{\bibfnamefont{F.}~\bibnamefont{Wu}},
  \bibinfo{author}{\bibfnamefont{A.~H.} \bibnamefont{MacDonald}},
  \bibnamefont{and} \bibinfo{author}{\bibfnamefont{I.}~\bibnamefont{Martin}},
  \bibinfo{journal}{Phys. Rev. Lett.} \textbf{\bibinfo{volume}{121}},
  \bibinfo{pages}{257001} (\bibinfo{year}{2018}).

\bibitem[{\citenamefont{Thomson et~al.}(2018)\citenamefont{Thomson, Chatterjee,
  Sachdev, and Scheurer}}]{PhysRevB.98.075109}
\bibinfo{author}{\bibfnamefont{A.}~\bibnamefont{Thomson}},
  \bibinfo{author}{\bibfnamefont{S.}~\bibnamefont{Chatterjee}},
  \bibinfo{author}{\bibfnamefont{S.}~\bibnamefont{Sachdev}}, \bibnamefont{and}
  \bibinfo{author}{\bibfnamefont{M.~S.} \bibnamefont{Scheurer}},
  \bibinfo{journal}{Phys. Rev. B} \textbf{\bibinfo{volume}{98}},
  \bibinfo{pages}{075109} (\bibinfo{year}{2018}).

\bibitem[{\citenamefont{Ochi et~al.}(2018)\citenamefont{Ochi, Koshino, and
  Kuroki}}]{OchPRB982018}
\bibinfo{author}{\bibfnamefont{M.}~\bibnamefont{Ochi}},
  \bibinfo{author}{\bibfnamefont{M.}~\bibnamefont{Koshino}}, \bibnamefont{and}
  \bibinfo{author}{\bibfnamefont{K.}~\bibnamefont{Kuroki}},
  \bibinfo{journal}{Phys. Rev. B} \textbf{\bibinfo{volume}{98}},
  \bibinfo{pages}{081102(R)} (\bibinfo{year}{2018}).

\bibitem[{\citenamefont{Guinea and Walet}(2018)}]{Guinea13174}
\bibinfo{author}{\bibfnamefont{F.}~\bibnamefont{Guinea}} \bibnamefont{and}
  \bibinfo{author}{\bibfnamefont{N.~R.} \bibnamefont{Walet}},
  \bibinfo{journal}{Proceedings of the National Academy of Sciences}
  \textbf{\bibinfo{volume}{115}}, \bibinfo{pages}{13174}
  (\bibinfo{year}{2018}).

\bibitem[{\citenamefont{{Po} et~al.}(2018)\citenamefont{{Po}, {Zou}, {Senthil},
  and {Vishwanath}}}]{2018arXiv180802482P}
\bibinfo{author}{\bibfnamefont{H.~C.} \bibnamefont{{Po}}},
  \bibinfo{author}{\bibfnamefont{L.}~\bibnamefont{{Zou}}},
  \bibinfo{author}{\bibfnamefont{T.}~\bibnamefont{{Senthil}}},
  \bibnamefont{and}
  \bibinfo{author}{\bibfnamefont{A.}~\bibnamefont{{Vishwanath}}},
  \bibinfo{journal}{arXiv e-prints}  (\bibinfo{year}{2018}),
  \eprint{1808.02482}.

\bibitem[{\citenamefont{Laksono et~al.}(2018)\citenamefont{Laksono, Leaw,
  Reaves, Singh, Wang, Adam, and Gu}}]{LAKSONO201838}
\bibinfo{author}{\bibfnamefont{E.}~\bibnamefont{Laksono}},
  \bibinfo{author}{\bibfnamefont{J.~N.} \bibnamefont{Leaw}},
  \bibinfo{author}{\bibfnamefont{A.}~\bibnamefont{Reaves}},
  \bibinfo{author}{\bibfnamefont{M.}~\bibnamefont{Singh}},
  \bibinfo{author}{\bibfnamefont{X.}~\bibnamefont{Wang}},
  \bibinfo{author}{\bibfnamefont{S.}~\bibnamefont{Adam}}, \bibnamefont{and}
  \bibinfo{author}{\bibfnamefont{X.}~\bibnamefont{Gu}}, \bibinfo{journal}{Solid
  State Communications} \textbf{\bibinfo{volume}{282}}, \bibinfo{pages}{38 }
  (\bibinfo{year}{2018}).

\bibitem[{\citenamefont{Venderbos and Fernandes}(2018)}]{VenPRB982018}
\bibinfo{author}{\bibfnamefont{J.~W.~F.} \bibnamefont{Venderbos}}
  \bibnamefont{and} \bibinfo{author}{\bibfnamefont{R.~M.}
  \bibnamefont{Fernandes}}, \bibinfo{journal}{Phys. Rev. B}
  \textbf{\bibinfo{volume}{98}}, \bibinfo{pages}{245103}
  (\bibinfo{year}{2018}).

\bibitem[{\citenamefont{Choi and Choi}(2018)}]{PhysRevB.98.241412}
\bibinfo{author}{\bibfnamefont{Y.~W.} \bibnamefont{Choi}} \bibnamefont{and}
  \bibinfo{author}{\bibfnamefont{H.~J.} \bibnamefont{Choi}},
  \bibinfo{journal}{Phys. Rev. B} \textbf{\bibinfo{volume}{98}},
  \bibinfo{pages}{241412(R)} (\bibinfo{year}{2018}).

\bibitem[{\citenamefont{{Kozii} et~al.}(2018)\citenamefont{{Kozii}, {Isobe},
  {Venderbos}, and {Fu}}}]{2018arXiv181004159K}
\bibinfo{author}{\bibfnamefont{V.}~\bibnamefont{{Kozii}}},
  \bibinfo{author}{\bibfnamefont{H.}~\bibnamefont{{Isobe}}},
  \bibinfo{author}{\bibfnamefont{J.~W.~F.} \bibnamefont{{Venderbos}}},
  \bibnamefont{and} \bibinfo{author}{\bibfnamefont{L.}~\bibnamefont{{Fu}}},
  \bibinfo{journal}{arXiv e-prints}  (\bibinfo{year}{2018}),
  \eprint{1810.04159}.

\bibitem[{\citenamefont{{Kang} and {Vafek}}(2018)}]{2018arXiv181008642K}
\bibinfo{author}{\bibfnamefont{J.}~\bibnamefont{{Kang}}} \bibnamefont{and}
  \bibinfo{author}{\bibfnamefont{O.}~\bibnamefont{{Vafek}}},
  \bibinfo{journal}{arXiv e-prints}  (\bibinfo{year}{2018}),
  \eprint{1810.08642}.

\bibitem[{\citenamefont{{Seo} et~al.}(2018)\citenamefont{{Seo}, {Kotov}, and
  {Uchoa}}}]{2018arXiv181202550S}
\bibinfo{author}{\bibfnamefont{K.}~\bibnamefont{{Seo}}},
  \bibinfo{author}{\bibfnamefont{V.~N.} \bibnamefont{{Kotov}}},
  \bibnamefont{and} \bibinfo{author}{\bibfnamefont{B.}~\bibnamefont{{Uchoa}}},
  \bibinfo{journal}{arXiv e-prints}  (\bibinfo{year}{2018}),
  \eprint{1812.02550}.

\bibitem[{\citenamefont{{Xie} and {MacDonald}}(2018)}]{2018arXiv181204213X}
\bibinfo{author}{\bibfnamefont{M.}~\bibnamefont{{Xie}}} \bibnamefont{and}
  \bibinfo{author}{\bibfnamefont{A.~H.} \bibnamefont{{MacDonald}}},
  \bibinfo{journal}{arXiv e-prints}  (\bibinfo{year}{2018}),
  \eprint{1812.04213}.

\bibitem[{\citenamefont{Wehling et~al.}(2011)\citenamefont{Wehling, \ifmmode
  \mbox{\c{S}}\else \c{S}\fi{}a\ifmmode \mbox{\c{s}}\else \c{s}\fi{}\ifmmode
  \imath \else \i \fi{}o\ifmmode~\breve{g}\else \u{g}\fi{}lu, Friedrich,
  Lichtenstein, Katsnelson, and Bl\"ugel}}]{WehPRL1062011}
\bibinfo{author}{\bibfnamefont{T.~O.} \bibnamefont{Wehling}},
  \bibinfo{author}{\bibfnamefont{E.}~\bibnamefont{\ifmmode \mbox{\c{S}}\else
  \c{S}\fi{}a\ifmmode \mbox{\c{s}}\else \c{s}\fi{}\ifmmode \imath \else \i
  \fi{}o\ifmmode~\breve{g}\else \u{g}\fi{}lu}},
  \bibinfo{author}{\bibfnamefont{C.}~\bibnamefont{Friedrich}},
  \bibinfo{author}{\bibfnamefont{A.~I.} \bibnamefont{Lichtenstein}},
  \bibinfo{author}{\bibfnamefont{M.~I.} \bibnamefont{Katsnelson}},
  \bibnamefont{and} \bibinfo{author}{\bibfnamefont{S.}~\bibnamefont{Bl\"ugel}},
  \bibinfo{journal}{Phys. Rev. Lett.} \textbf{\bibinfo{volume}{106}},
  \bibinfo{pages}{236805} (\bibinfo{year}{2011}).

\bibitem[{\citenamefont{R\"osner et~al.}(2015)\citenamefont{R\"osner, \ifmmode
  \mbox{\c{S}}\else \c{S}\fi{}a\ifmmode \mbox{\c{s}}\else \c{s}\fi{}\ifmmode
  \imath \else \i \fi{}o\ifmmode~\breve{g}\else \u{g}\fi{}lu, Friedrich,
  Bl\"ugel, and Wehling}}]{RosPRB922015}
\bibinfo{author}{\bibfnamefont{M.}~\bibnamefont{R\"osner}},
  \bibinfo{author}{\bibfnamefont{E.}~\bibnamefont{\ifmmode \mbox{\c{S}}\else
  \c{S}\fi{}a\ifmmode \mbox{\c{s}}\else \c{s}\fi{}\ifmmode \imath \else \i
  \fi{}o\ifmmode~\breve{g}\else \u{g}\fi{}lu}},
  \bibinfo{author}{\bibfnamefont{C.}~\bibnamefont{Friedrich}},
  \bibinfo{author}{\bibfnamefont{S.}~\bibnamefont{Bl\"ugel}}, \bibnamefont{and}
  \bibinfo{author}{\bibfnamefont{T.~O.} \bibnamefont{Wehling}},
  \bibinfo{journal}{Phys. Rev. B} \textbf{\bibinfo{volume}{92}},
  \bibinfo{pages}{085102} (\bibinfo{year}{2015}).

\bibitem[{\citenamefont{Sch\"onhoff et~al.}(2016)\citenamefont{Sch\"onhoff,
  R\"osner, Groenewald, Haas, and Wehling}}]{schonhoff_interplay_2016}
\bibinfo{author}{\bibfnamefont{G.}~\bibnamefont{Sch\"onhoff}},
  \bibinfo{author}{\bibfnamefont{M.}~\bibnamefont{R\"osner}},
  \bibinfo{author}{\bibfnamefont{R.~E.} \bibnamefont{Groenewald}},
  \bibinfo{author}{\bibfnamefont{S.}~\bibnamefont{Haas}}, \bibnamefont{and}
  \bibinfo{author}{\bibfnamefont{T.~O.} \bibnamefont{Wehling}},
  \bibinfo{journal}{Phys. Rev. B} \textbf{\bibinfo{volume}{94}},
  \bibinfo{pages}{134504} (\bibinfo{year}{2016}).

\bibitem[{\citenamefont{Stauber and Kohler}(2016)}]{tobias2016}
\bibinfo{author}{\bibfnamefont{T.}~\bibnamefont{Stauber}} \bibnamefont{and}
  \bibinfo{author}{\bibfnamefont{H.}~\bibnamefont{Kohler}},
  \bibinfo{journal}{Nano Letters} \textbf{\bibinfo{volume}{16}},
  \bibinfo{pages}{6844} (\bibinfo{year}{2016}).

\bibitem[{\citenamefont{{Zhang}}(2018)}]{2018arXiv180409047Z}
\bibinfo{author}{\bibfnamefont{L.}~\bibnamefont{{Zhang}}},
  \bibinfo{journal}{arXiv e-prints}  (\bibinfo{year}{2018}),
  \eprint{1804.09047}.

\bibitem[{\citenamefont{Gonz\'alez and Stauber}(2019)}]{PhysRevLett.122.026801}
\bibinfo{author}{\bibfnamefont{J.}~\bibnamefont{Gonz\'alez}} \bibnamefont{and}
  \bibinfo{author}{\bibfnamefont{T.}~\bibnamefont{Stauber}},
  \bibinfo{journal}{Phys. Rev. Lett.} \textbf{\bibinfo{volume}{122}},
  \bibinfo{pages}{026801} (\bibinfo{year}{2019}).

\bibitem[{\citenamefont{{Sboychakov} et~al.}(2018)\citenamefont{{Sboychakov},
  {Rozhkov}, {Rakhmanov}, and {Nori}}}]{2018arXiv180708190S}
\bibinfo{author}{\bibfnamefont{A.~O.} \bibnamefont{{Sboychakov}}},
  \bibinfo{author}{\bibfnamefont{A.~V.} \bibnamefont{{Rozhkov}}},
  \bibinfo{author}{\bibfnamefont{A.~L.} \bibnamefont{{Rakhmanov}}},
  \bibnamefont{and} \bibinfo{author}{\bibfnamefont{F.}~\bibnamefont{{Nori}}},
  \bibinfo{journal}{arXiv e-prints}  (\bibinfo{year}{2018}),
  \eprint{1807.08190}.

\bibitem[{\citenamefont{{Fu} et~al.}(2018)\citenamefont{{Fu}, {K{\"o}nig},
  {Wilson}, {Chou}, and {Pixley}}}]{2018arXiv180904604F}
\bibinfo{author}{\bibfnamefont{Y.}~\bibnamefont{{Fu}}},
  \bibinfo{author}{\bibfnamefont{E.~J.} \bibnamefont{{K{\"o}nig}}},
  \bibinfo{author}{\bibfnamefont{J.~H.} \bibnamefont{{Wilson}}},
  \bibinfo{author}{\bibfnamefont{Y.-Z.} \bibnamefont{{Chou}}},
  \bibnamefont{and} \bibinfo{author}{\bibfnamefont{J.~H.}
  \bibnamefont{{Pixley}}}, \bibinfo{journal}{arXiv e-prints}
  (\bibinfo{year}{2018}), \eprint{1809.04604}.

\bibitem[{\citenamefont{Gonz\'alez et~al.}(1999)\citenamefont{Gonz\'alez,
  Guinea, and Vozmediano}}]{GonPRB591999}
\bibinfo{author}{\bibfnamefont{J.}~\bibnamefont{Gonz\'alez}},
  \bibinfo{author}{\bibfnamefont{F.}~\bibnamefont{Guinea}}, \bibnamefont{and}
  \bibinfo{author}{\bibfnamefont{M.~A.~H.} \bibnamefont{Vozmediano}},
  \bibinfo{journal}{Phys. Rev. B} \textbf{\bibinfo{volume}{59}},
  \bibinfo{pages}{R2474} (\bibinfo{year}{1999}).

\bibitem[{SM()}]{SM}
\bibinfo{note}{See Supplemental Material for details of the low-energy
  continuum model of the electronic bands in MA-tBLG, for an analysis of the
  matrix structure of polarization operators, for a study of the dielectric
  function $\varepsilon (q)$ in different dielectric environments, for a
  discussion of the temperature dependence of the polarization functions, and
  for investigations of the dependencies of internal screening on twist angles,
  doping, interlayer coupling parameters and energy gaps in the electronic
  excitation spectrum.}

\bibitem[{\citenamefont{Vonsovsky and Katsnelson}(1989)}]{vonkat_1989}
\bibinfo{author}{\bibfnamefont{S.~V.} \bibnamefont{Vonsovsky}}
  \bibnamefont{and} \bibinfo{author}{\bibfnamefont{M.~I.}
  \bibnamefont{Katsnelson}}, \emph{\bibinfo{title}{Quantum Solid-State
  Physics}} (\bibinfo{publisher}{Springer-Verlag Berlin Heidelberg},
  \bibinfo{year}{1989}).

\bibitem[{\citenamefont{Aryasetiawan et~al.}(2004)\citenamefont{Aryasetiawan,
  Imada, Georges, Kotliar, Biermann, and Lichtenstein}}]{PhysRevB.70.195104}
\bibinfo{author}{\bibfnamefont{F.}~\bibnamefont{Aryasetiawan}},
  \bibinfo{author}{\bibfnamefont{M.}~\bibnamefont{Imada}},
  \bibinfo{author}{\bibfnamefont{A.}~\bibnamefont{Georges}},
  \bibinfo{author}{\bibfnamefont{G.}~\bibnamefont{Kotliar}},
  \bibinfo{author}{\bibfnamefont{S.}~\bibnamefont{Biermann}}, \bibnamefont{and}
  \bibinfo{author}{\bibfnamefont{A.~I.} \bibnamefont{Lichtenstein}},
  \bibinfo{journal}{Phys. Rev. B} \textbf{\bibinfo{volume}{70}},
  \bibinfo{pages}{195104} (\bibinfo{year}{2004}).

\bibitem[{\citenamefont{van Loon et~al.}(2014)\citenamefont{van Loon,
  Hafermann, Lichtenstein, Rubtsov, and Katsnelson}}]{vanLoon_MottScreen}
\bibinfo{author}{\bibfnamefont{E.}~\bibnamefont{van Loon}},
  \bibinfo{author}{\bibfnamefont{H.}~\bibnamefont{Hafermann}},
  \bibinfo{author}{\bibfnamefont{A.}~\bibnamefont{Lichtenstein}},
  \bibinfo{author}{\bibfnamefont{A.}~\bibnamefont{Rubtsov}}, \bibnamefont{and}
  \bibinfo{author}{\bibfnamefont{M.}~\bibnamefont{Katsnelson}},
  \bibinfo{journal}{Phys. Rev. Lett.} \textbf{\bibinfo{volume}{113}},
  \bibinfo{pages}{246407} (\bibinfo{year}{2014}).

\bibitem[{Mot()}]{MottScreen}
\bibinfo{note}{Fig. 3 of Ref. \cite{vanLoon_MottScreen} shows that the spectral
  weight in the loss function ${\rm Im} 1/\epsilon$ is suppressed when going
  into a Mott state, particularly in the important long-wavelength and low
  energy region. This suppression of spectral weight implies that screening by
  the correlated bands is strongly suppressed in a Mott insulating state.}

\bibitem[{GGp()}]{GGpol}
\bibinfo{note}{A study of the full matrix structure of
  $\Pi_0^{\bold{G},\bold{G}'}$ is given in the Supplemental Material \cite{SM}.
  The diagonal matrix elements of $\Pi_0^{\bold{G},\bold{G}'}$ are by far the
  largest at all $\bold{q}$ in cRPA and also leading in RPA except for certain
  $\bold{q}$ close to the mini BZ boundaries. Here, we analyze the diagonal
  matrix elements $\bold{G}=\bold{G}'$ by considering the single element $\Pi_0
  \equiv \Pi_0^{\bold{G}=\bold{G}' = \bold{0}}$ and allowing for $\bold{q}$ to
  run out of the first mini-BZ in the sense of an extended zone scheme.}

\bibitem[{\citenamefont{Katsnelson}(2012)}]{katsnelson_2012}
\bibinfo{author}{\bibfnamefont{M.~I.} \bibnamefont{Katsnelson}},
  \emph{\bibinfo{title}{Graphene: Carbon in Two Dimensions}}
  (\bibinfo{publisher}{Cambridge University Press}, \bibinfo{year}{2012}).

\bibitem[{\citenamefont{Yang et~al.}(1991)\citenamefont{Yang, Guo, Chan, Wong,
  and Ching}}]{PhysRevA.43.1186}
\bibinfo{author}{\bibfnamefont{X.~L.} \bibnamefont{Yang}},
  \bibinfo{author}{\bibfnamefont{S.~H.} \bibnamefont{Guo}},
  \bibinfo{author}{\bibfnamefont{F.~T.} \bibnamefont{Chan}},
  \bibinfo{author}{\bibfnamefont{K.~W.} \bibnamefont{Wong}}, \bibnamefont{and}
  \bibinfo{author}{\bibfnamefont{W.~Y.} \bibnamefont{Ching}},
  \bibinfo{journal}{Phys. Rev. A} \textbf{\bibinfo{volume}{43}},
  \bibinfo{pages}{1186} (\bibinfo{year}{1991}).

\bibitem[{\citenamefont{Hwang and Das~Sarma}(2008)}]{HwaPRL1012008}
\bibinfo{author}{\bibfnamefont{E.~H.} \bibnamefont{Hwang}} \bibnamefont{and}
  \bibinfo{author}{\bibfnamefont{S.}~\bibnamefont{Das~Sarma}},
  \bibinfo{journal}{Phys. Rev. Lett.} \textbf{\bibinfo{volume}{101}},
  \bibinfo{pages}{156802} (\bibinfo{year}{2008}).

\bibitem[{\citenamefont{Tarnopolsky et~al.}(2019)\citenamefont{Tarnopolsky,
  Kruchkov, and Vishwanath}}]{tarnopolsky_origin_2019}
\bibinfo{author}{\bibfnamefont{G.}~\bibnamefont{Tarnopolsky}},
  \bibinfo{author}{\bibfnamefont{A.~J.} \bibnamefont{Kruchkov}},
  \bibnamefont{and}
  \bibinfo{author}{\bibfnamefont{A.}~\bibnamefont{Vishwanath}},
  \bibinfo{journal}{Phys. Rev. Lett.} \textbf{\bibinfo{volume}{122}},
  \bibinfo{pages}{106405} (\bibinfo{year}{2019}).

\bibitem[{\citenamefont{Lu et~al.}(2019)\citenamefont{Lu, Stepanov, Yang, Xie,
  Aamir, Das, Urgell, Watanabe, Taniguchi, Zhang
  et~al.}}]{lu_superconductors_2019}
\bibinfo{author}{\bibfnamefont{X.}~\bibnamefont{Lu}},
  \bibinfo{author}{\bibfnamefont{P.}~\bibnamefont{Stepanov}},
  \bibinfo{author}{\bibfnamefont{W.}~\bibnamefont{Yang}},
  \bibinfo{author}{\bibfnamefont{M.}~\bibnamefont{Xie}},
  \bibinfo{author}{\bibfnamefont{M.~A.} \bibnamefont{Aamir}},
  \bibinfo{author}{\bibfnamefont{I.}~\bibnamefont{Das}},
  \bibinfo{author}{\bibfnamefont{C.}~\bibnamefont{Urgell}},
  \bibinfo{author}{\bibfnamefont{K.}~\bibnamefont{Watanabe}},
  \bibinfo{author}{\bibfnamefont{T.}~\bibnamefont{Taniguchi}},
  \bibinfo{author}{\bibfnamefont{G.}~\bibnamefont{Zhang}}, \bibnamefont{et~al.}
  (\bibinfo{year}{2019}), \bibinfo{note}{arXiv: 1903.06513}.

\end{thebibliography}

\newpage
$\ $

\newpage

\beginsupplement{
\section{SUPPLEMENTAL MATERIAL}

In section A of this Supplemental Material, we give a short description of the low-energy continuum model developed in Refs. \cite{BisPNAS1082011,KosPRX82018} for tBLG and employed here. We show the band structure and the DOS around the charge neutrality point (CNP) for the first magic angle $1.05 \: ^{\circ}$. In section B, we discuss the matrix structure of the polarization operator given in Eq. (1) of the main text as obtained in RPA and cRPA. Section C analyzes the dielectric function $\varepsilon (q)$ of MA-tBLG in different dielectric environments. Section D discusses the temperature dependence of the polarization functions in RPA and cRPA. In section E, we study the dependencies of internal screening in tBLG on twist angle, doping and variations in vertical coupling parameters used in the continuum model. In section F, we investigate how gaps on the order of $0.1$~meV to $1$~meV opened at the CNP and at half-filling of the flat bands affect the polarization function of MA-tBLG.

\subsection{LOW-ENERGY CONTINUUM MODEL FOR tBLG}

\begin{figure*}
\leavevmode
\includegraphics[clip,width=0.49\textwidth]{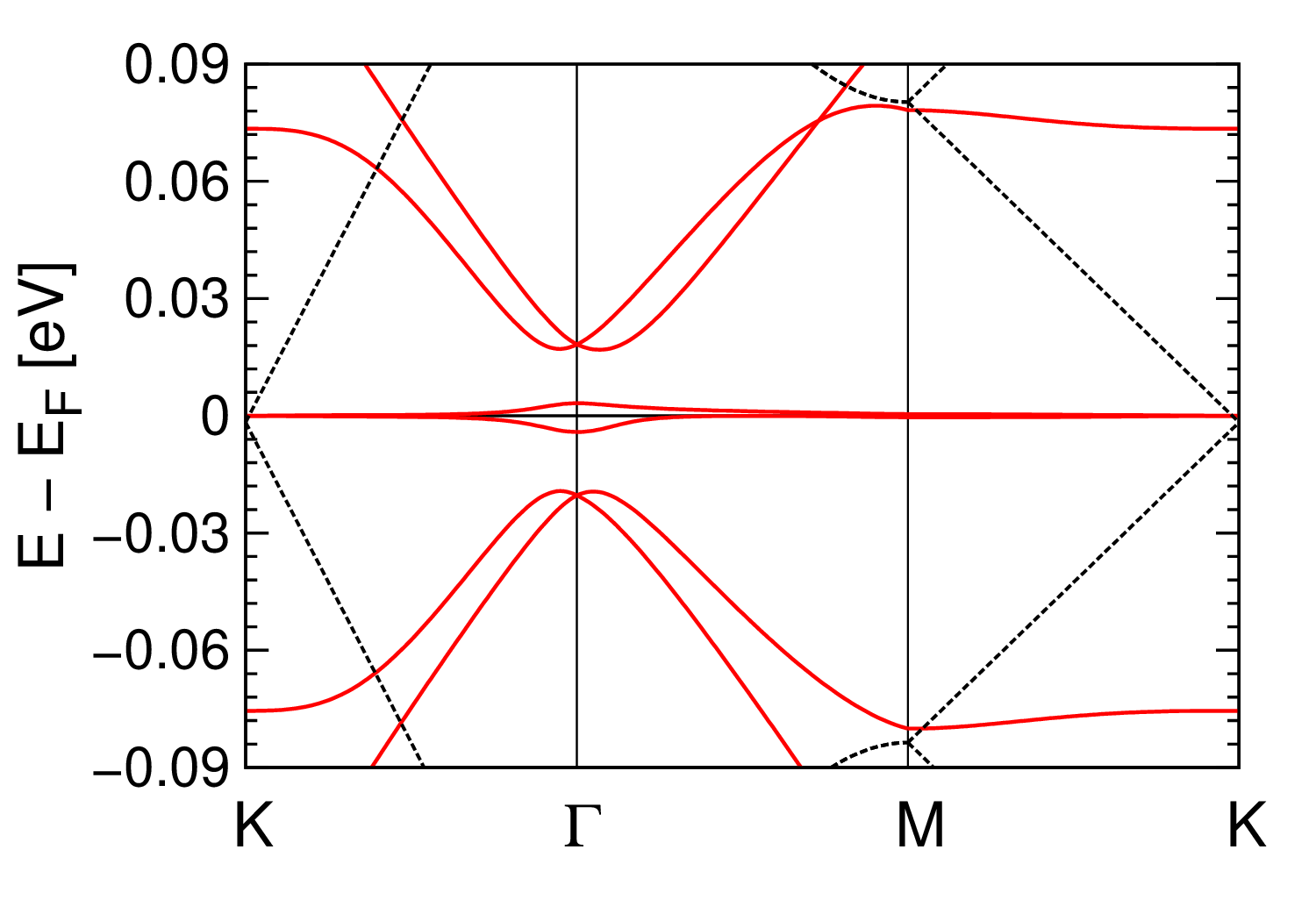}
\includegraphics[clip,width=0.49\textwidth]{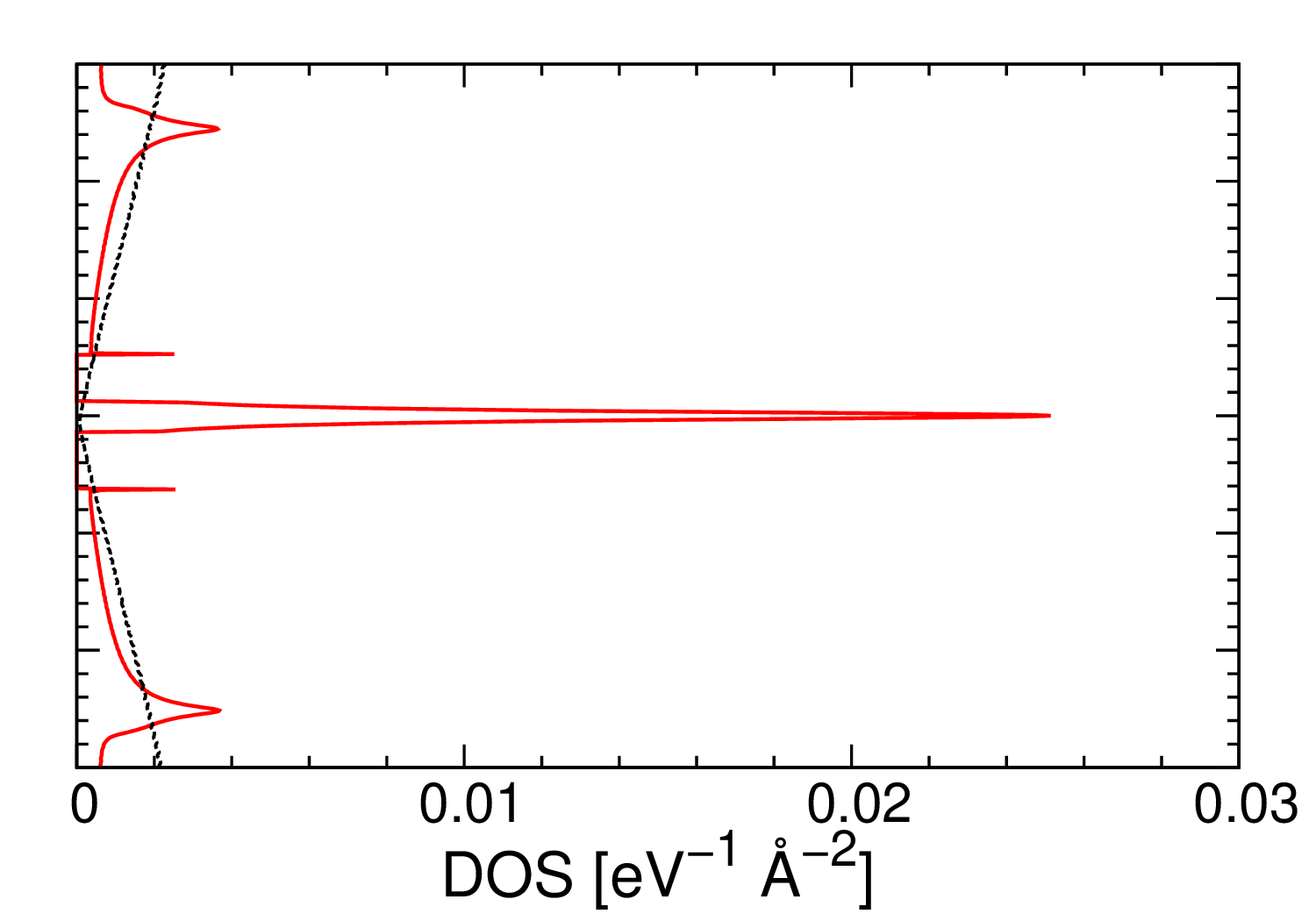}
\caption{(Color online) (a) Band structure and (b) density of states (DOS) at the first magic angle $1.05 \: ^{\circ}$ for valley $\xi=+$ for MA-tBLG (red) as derived from the low-energy continuum model, and for uncoupled tBLG (dashed black). The DOS shown here is per valley and spin degrees of freedom.} 
\label{figSM1} 
\vspace{-0.3cm}
\end{figure*}

We outline here the low-energy continuum model which we used for the description of the electronic band structure of MA-tBLG. The model has been introduced in \cite{BisPNAS1082011,KosPRX82018}. By neglecting the intervalley coupling, the total low-energy Hamiltonian can be written as
\begin{equation}
H \approx \begin{pmatrix}
H^{+}	& 	0 \\
0	& 	H^{-} \\
\end{pmatrix},
\label{eqSM1}
\end{equation}
where $H^{\xi}$, $\xi=\pm$, describe the two valleys separately. The intravalley blocks are written in the space of the sublattices $(A,B)$ and layer $(1,2)$ degrees of freedom, $(A_1,B_1,A_2,B_2)^\xi$
\begin{equation}
H^{\xi} = \begin{pmatrix}
h_1	& 	h_{\theta}^{\dagger} \\
h_{\theta}	& 	h_2 \\
\end{pmatrix},
\label{eqSM2}
\end{equation}
where $h_l$ with $l=1,2$ are the intralayer terms, and $h_{\theta}$ the interlayer ones. Then, following derivations in \cite{KosPRX82018}, the intralayer term can be written as
\begin{equation}
h_{l} = - \hbar v_F \left[ \mathcal{R}(\pm \theta/2) (\bold{k} - \bold{K}^{\xi}_l) \right] \cdot (\xi \sigma_x, \sigma_y),
\label{eqSM2a}
\end{equation}
where $\pm$ stands for $l=1$ and $2$, respectively, $\hbar v_F / a = 2.1354 \: \text{eV}$, $\mathcal{R}(\theta)$ is the 2D rotation matrix, $\sigma_x$ and $\sigma_y$ are the 2D Pauli matrices and $\bold{K}_l^\xi$ are the rotated graphene $K$ points of each layer $l$ in each valley $\xi$. For the interlayer term we use
\begin{equation}
\begin{aligned}
h_{\theta} = 
& \begin{pmatrix}
u	& 	u' \\
u'	& 	u \\
\end{pmatrix} \\
& + \begin{pmatrix}
u	& 	u'\omega^{-\xi} \\
u'\omega^{\xi}	& 	u \\
\end{pmatrix} e^{i\xi \bold{G}_1^M \cdot \bold{r}} \\
& + \begin{pmatrix}
u	& 	u'\omega^{\xi} \\
u' \omega^{-\xi}	& 	u \\
\end{pmatrix} e^{i\xi (\bold{G}_1^M+\bold{G}_2^M) \cdot \bold{r}},
\end{aligned}
\label{eqSM2b}
\end{equation}
with $\omega=e^{i2\pi/3}$, the AA interlayer coupling $u=0.0797 \: \text{eV}$ and the AB interlayer coupling $u'=0.0975 \: \text{eV}$. The inequality $u\neq u'$ accounts for possible corrugation effects in MA-tBLG \cite{KosPRX82018}. Uncoupled tBLG calculations are performed by switching off the interlayer terms, i.e. $h_\theta \equiv 0$. Due to the valley and spin degeneracy, we simply diagonalize $H^+$ to calculate the intrinsic static polarization $\Pi_0^{\bold{G}, \bold{G}'}$ and including degeneracy factors $g_v=2$ and $g_s=2$ in Eq. (1).

The diagonalization of $H^+$ is performed in the $k$-space. Then, there is a coupling between Dirac electron states at each $\bold{k}$ vector in one layer with states at $\bold{k}+\bold{G}$ in the other layer, where $\bold{G}=m\bold{G}_1^M+n\bold{G}_2^M$ is the reciprocal vector associated with the moir\'e superlattice, $m$ and $n$ are integers, and $\bold{G}_1^M$ and $\bold{G}_2^M$ span the reciprocal lattice of tBLG. The electronic states are
\begin{equation}
\phi_{\bold{k}}^\alpha (\bold{r}) = \sum_{\bold{G}} c_{\bold{k}}^{\alpha} e^{i(\bold{k}+\bold{G})\cdot \bold{r}},
\label{eqSM3}
\end{equation}
where $\alpha$ encodes the sublattice / layer $X=A_1,B_1,A_2,B_2$ and band $n$ indices \cite{KosPRX82018}. These eigenstates are expanded up to a certain plane wave cutoff $G_c$. In this work, we use $G_c=8$ for the calculation of $\Pi_0 (q)$ in Fig. 2, and $G_c=2$ in the calculations of the matrices $\Pi_0^{\bold{G},\bold{G}'} (\bold{q})$ in Fig. \ref{figSM2}.

In Fig. \ref{figSM1}, we show the band structure and the DOS of MA-tBLG for the first magic angle $1.05 \: ^{\circ}$, as well as its comparison with uncoupled tBLG. The flat bands with bandwidth $W \approx 7 \: \text{meV}$, as well as the higher energy bands separated by gaps of the order of $10 \: \text{meV}$ from the flat bands are well reproduced. 

In cRPA, polarization processes taking place inside the flat band manifold are excluded. However, gapped transitions between states associated with peaks in the DOS, e.g. between the flat bands and higher energy DOS peaks around, are still possible. These gapped transitions are reminiscent of the atomic systems such as the hydrogen atom. The q-dependent polarizability from transitions between hydrogen $s$ and $p$-orbitals reads \cite{PhysRevA.43.1186}
\begin{equation}
\Pi_0^H (q) \sim \frac{q^2}{\left( 1 + \left( \frac{q}{b} \right)^2 \right)^5},
\label{eqSM3a}
\end{equation}
where $b$ is an effective inverse orbital radius. Superimposing this quasi-atomic model with a Dirac electron model capturing all higher energy processes leads to the ansatz of Eq. (4) from the main text. As shown in the main text, this ansatz yields a very good fit of the polarization function in the cRPA case.

\subsection{POLARIZATION MATRIX}

\begin{figure*}
\leavevmode
\includegraphics[clip,width=0.325\textwidth]{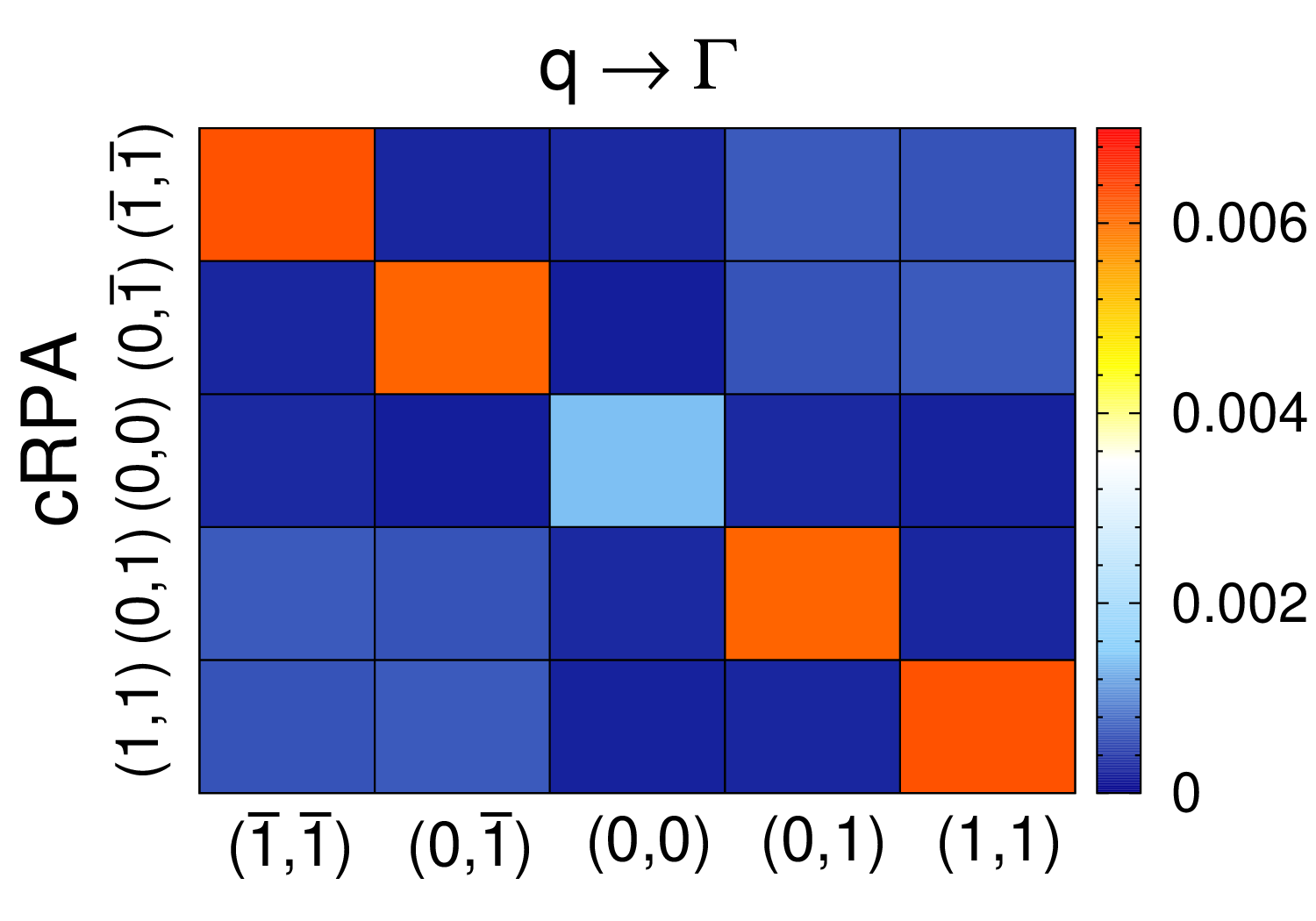}
\includegraphics[clip,width=0.325\textwidth]{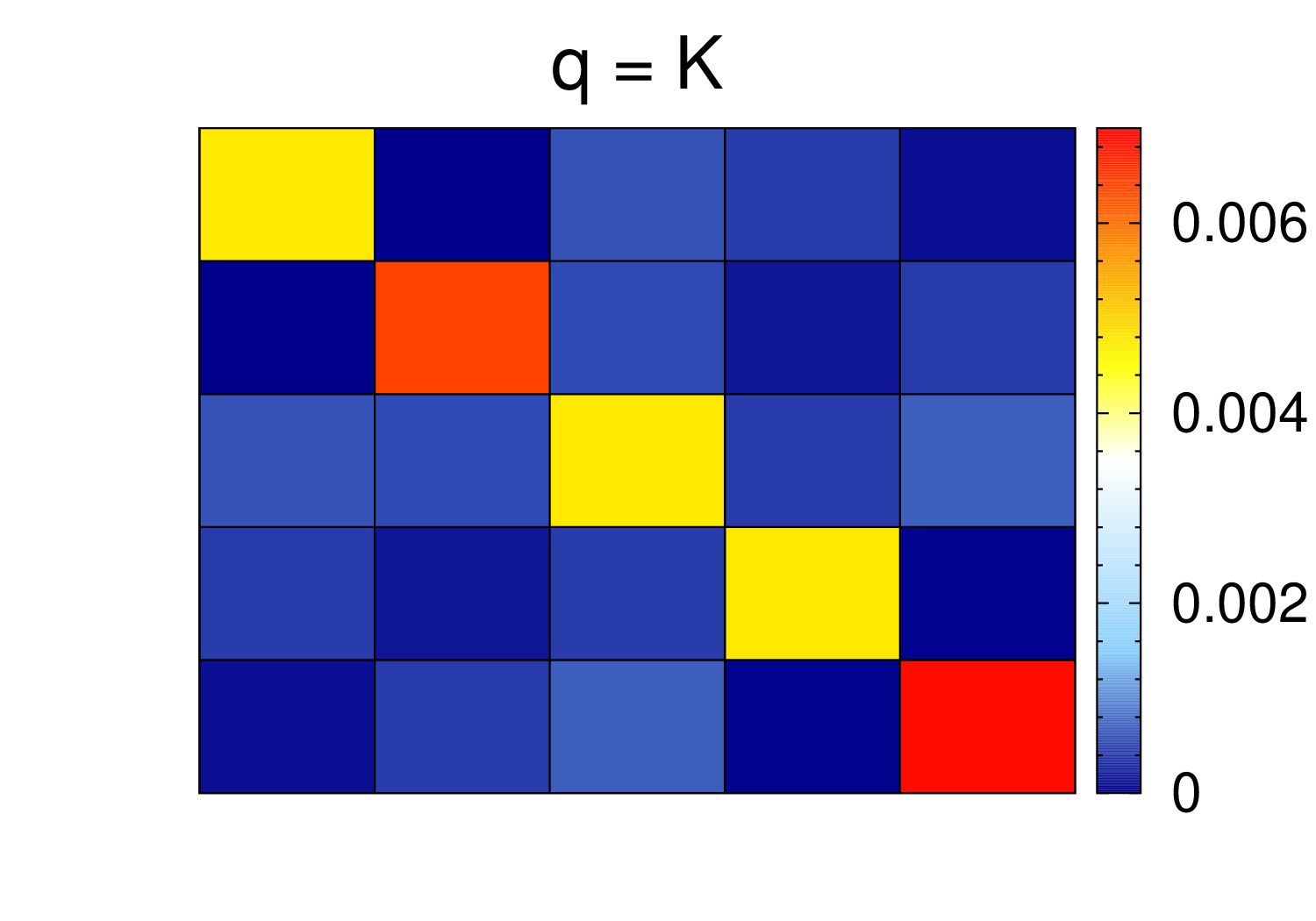} 
\includegraphics[clip,width=0.325\textwidth]{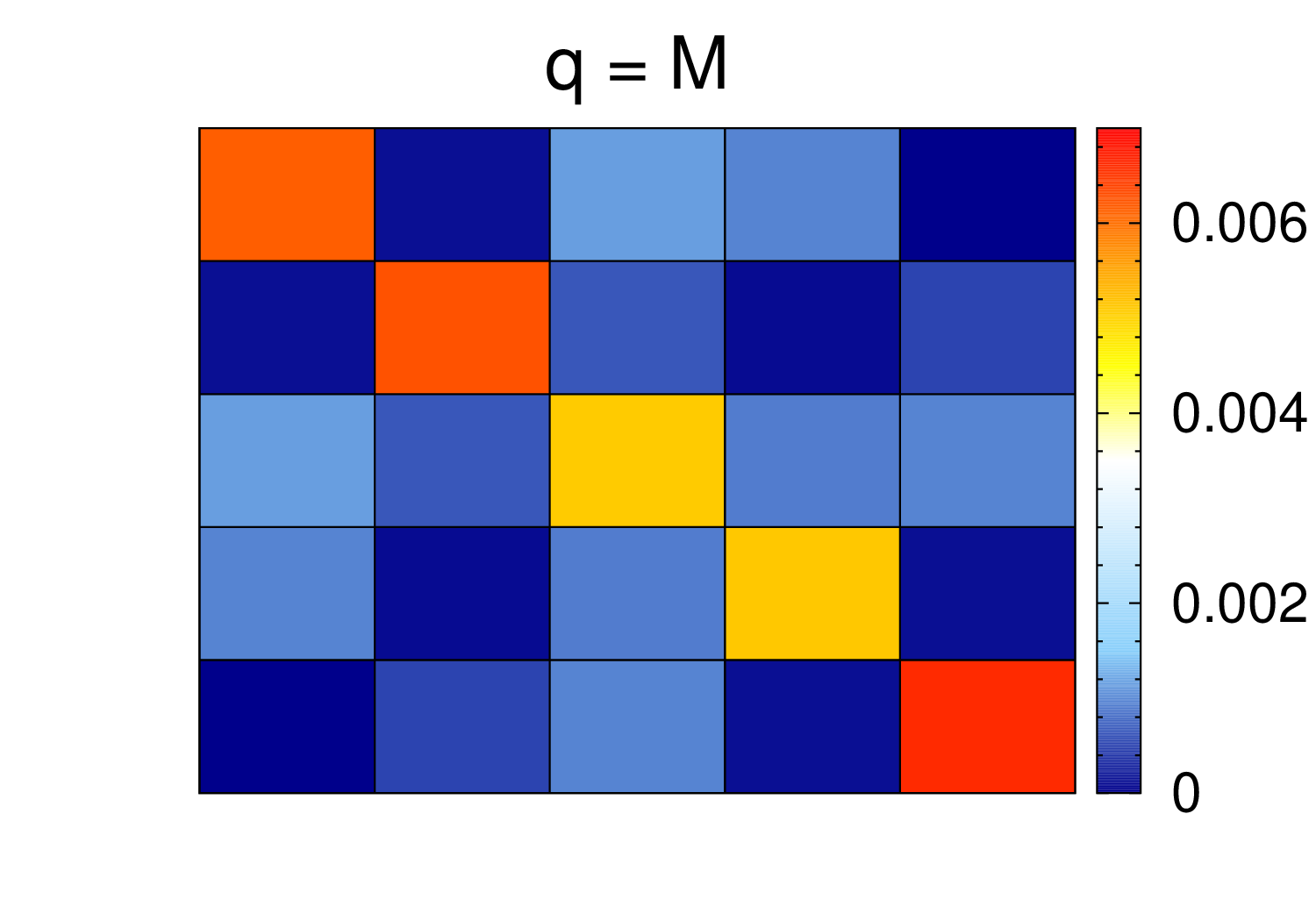} 
\includegraphics[clip,width=0.325\textwidth]{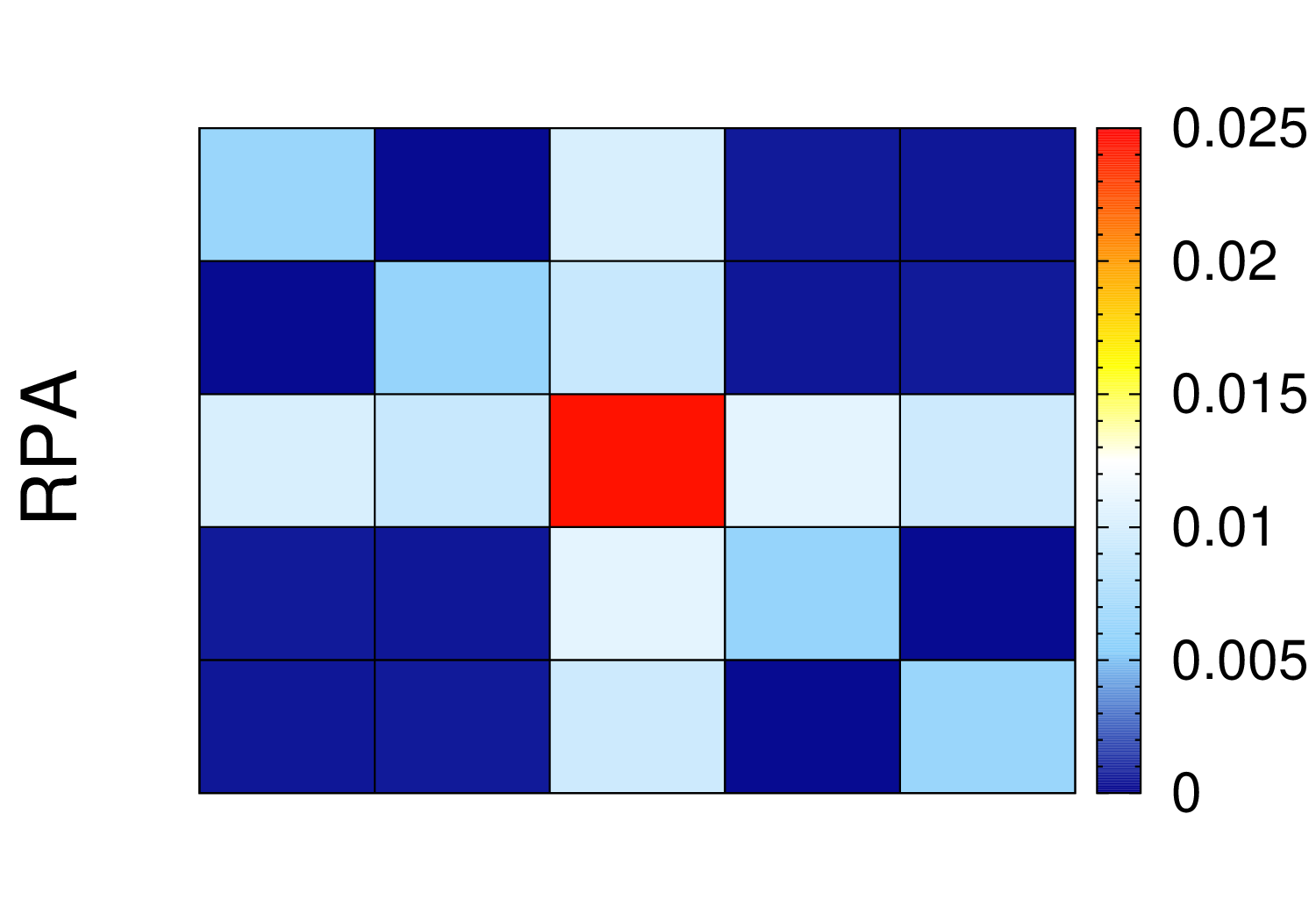}
\includegraphics[clip,width=0.325\textwidth]{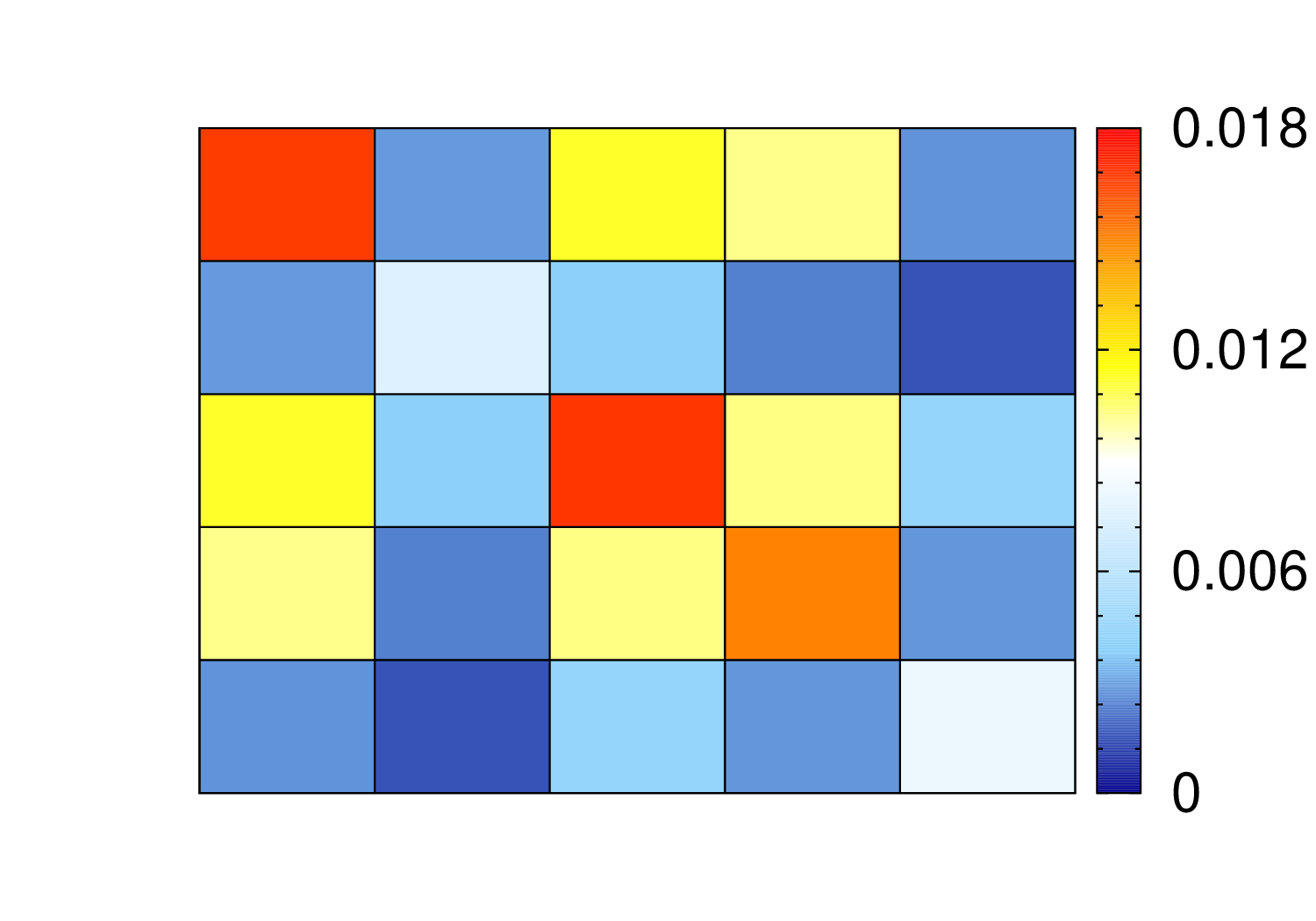} 
\includegraphics[clip,width=0.325\textwidth]{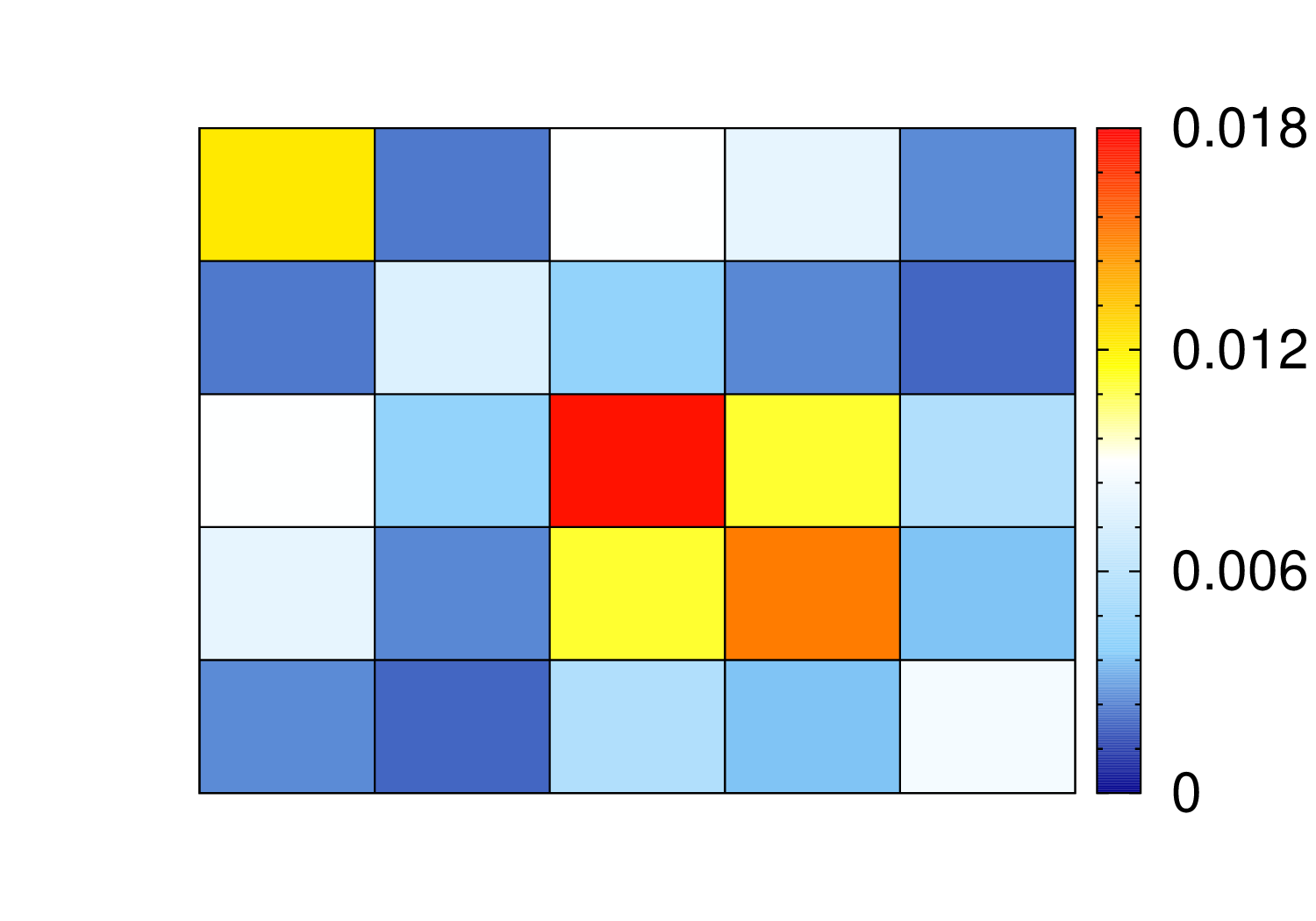} 
\caption{(Color online) Color maps of the cRPA (top row) and RPA (bottom row) intrinsic polarization matrices at $\bold{q} \rightarrow \bold{\Gamma}$ (left column), $\bold{K}$ (middle column) and $\bold{M}$ (right column) scattering vectors. Each matrix element corresponds to a pair of $\bold{G}$ and $\bold{G}'$ vectors, which are labelled according to the indices $(m,n)$, $\bold{G}=m\bold{G}_1^M+n\bold{G}_2^M$, and $\bar{1} \equiv -1$.} 
\label{figSM2} 
\vspace{-0.3cm}
\end{figure*}

Fig. \ref{figSM2} shows the polarization matrix $\Pi_0^{\bold{G},\bold{G}'}(\bold{q})$ for $\bold{q} \rightarrow \bold{\Gamma}$, $\bold{K}$ and $\bold{M}$ as obtained in RPA and cRPA using $G_c=2$.
In the cRPA case (top row of Fig. \ref{figSM2}), the diagonal elements are the leading elements for all $\bold{q}$. In the RPA case (bottom row of Fig. \ref{figSM2}), local field effects, i.e. off-diagonal matrix elements of $\Pi_0^{\bold{G},\bold{G}'}(\bold{q})$ with $\bold{G}\neq\bold{G}'$, become progresively more important when $\bold{q}$ is close to the mini BZ boundaries. However, the diagonal elements are still largest for small $q$, which is the decisive region of $q$-space to determine the impact of substrate screening effects.

\subsection{DIELECTRIC FUNCTION FOR tBLG}


\begin{figure*}
\leavevmode
\includegraphics[clip,width=0.325\textwidth]{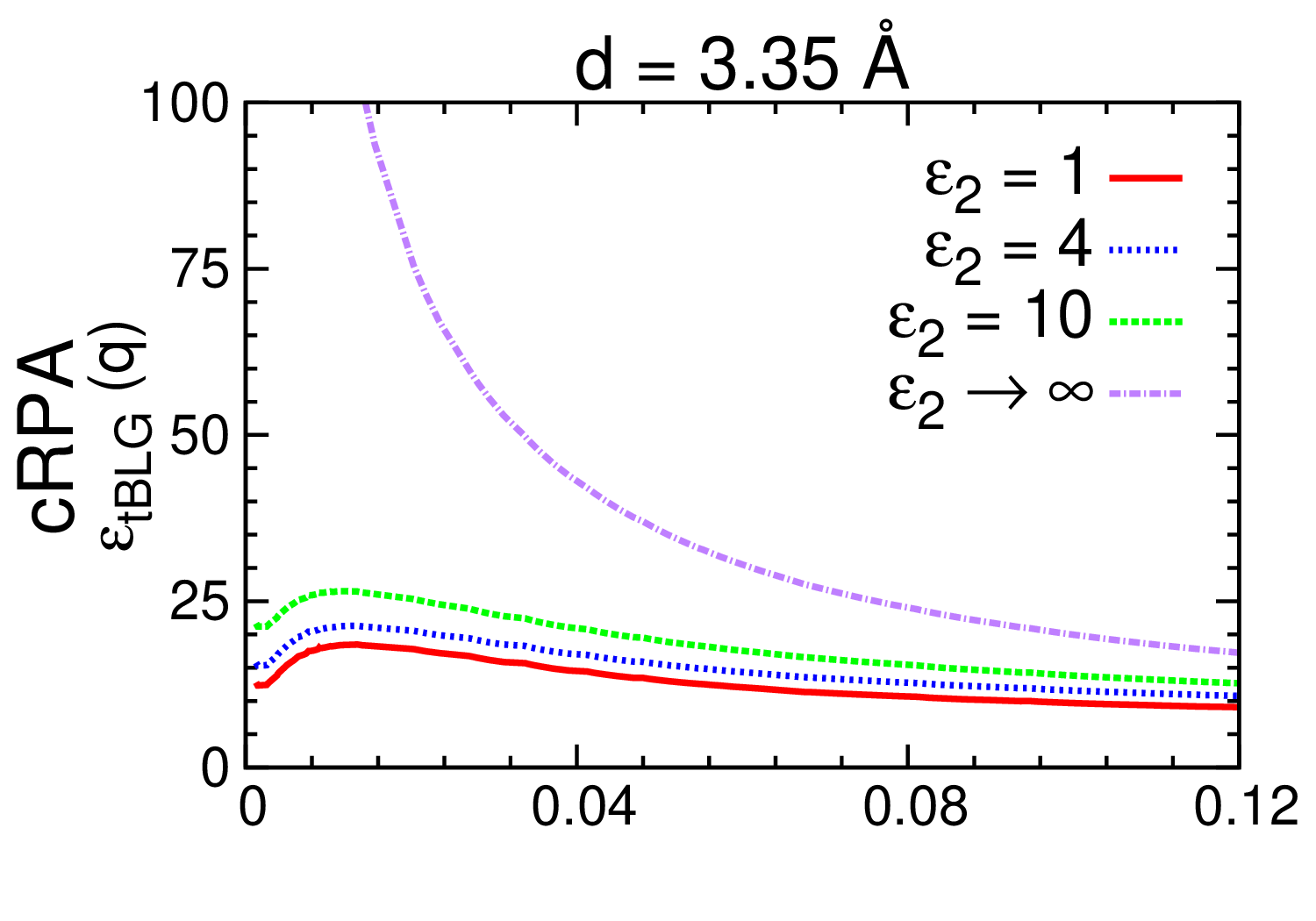}
\includegraphics[clip,width=0.325\textwidth]{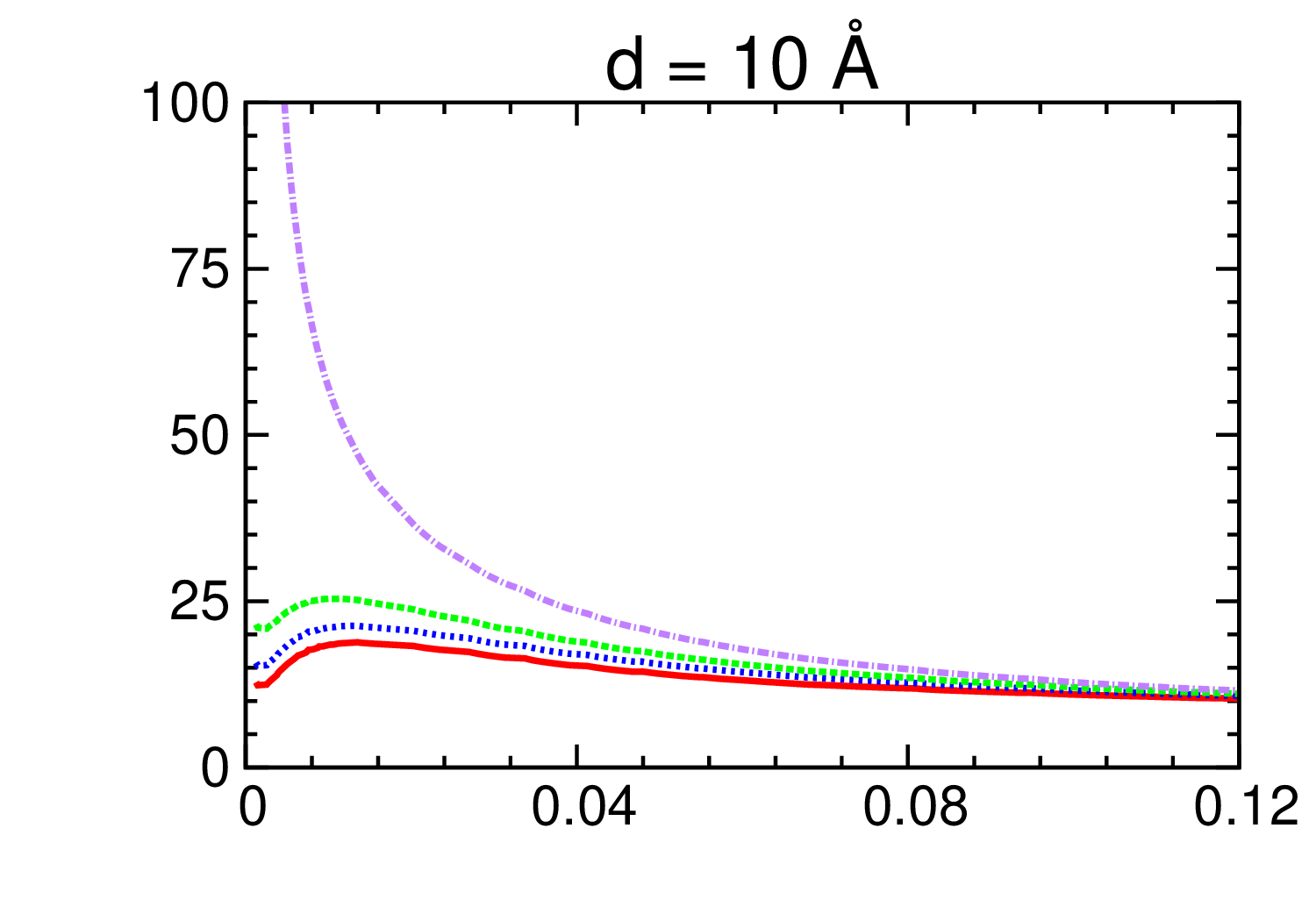} 
\includegraphics[clip,width=0.325\textwidth]{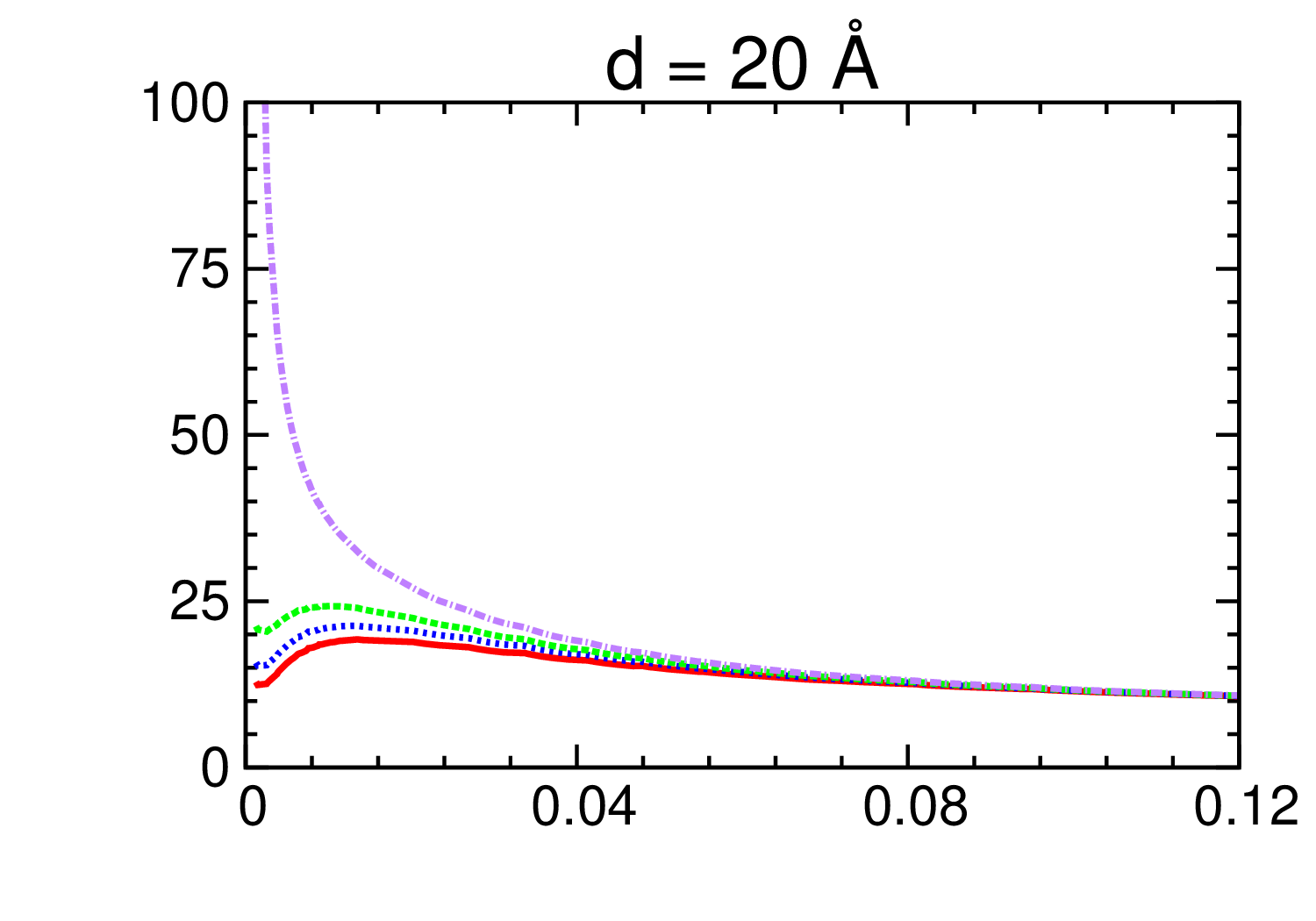}
\includegraphics[clip,width=0.325\textwidth]{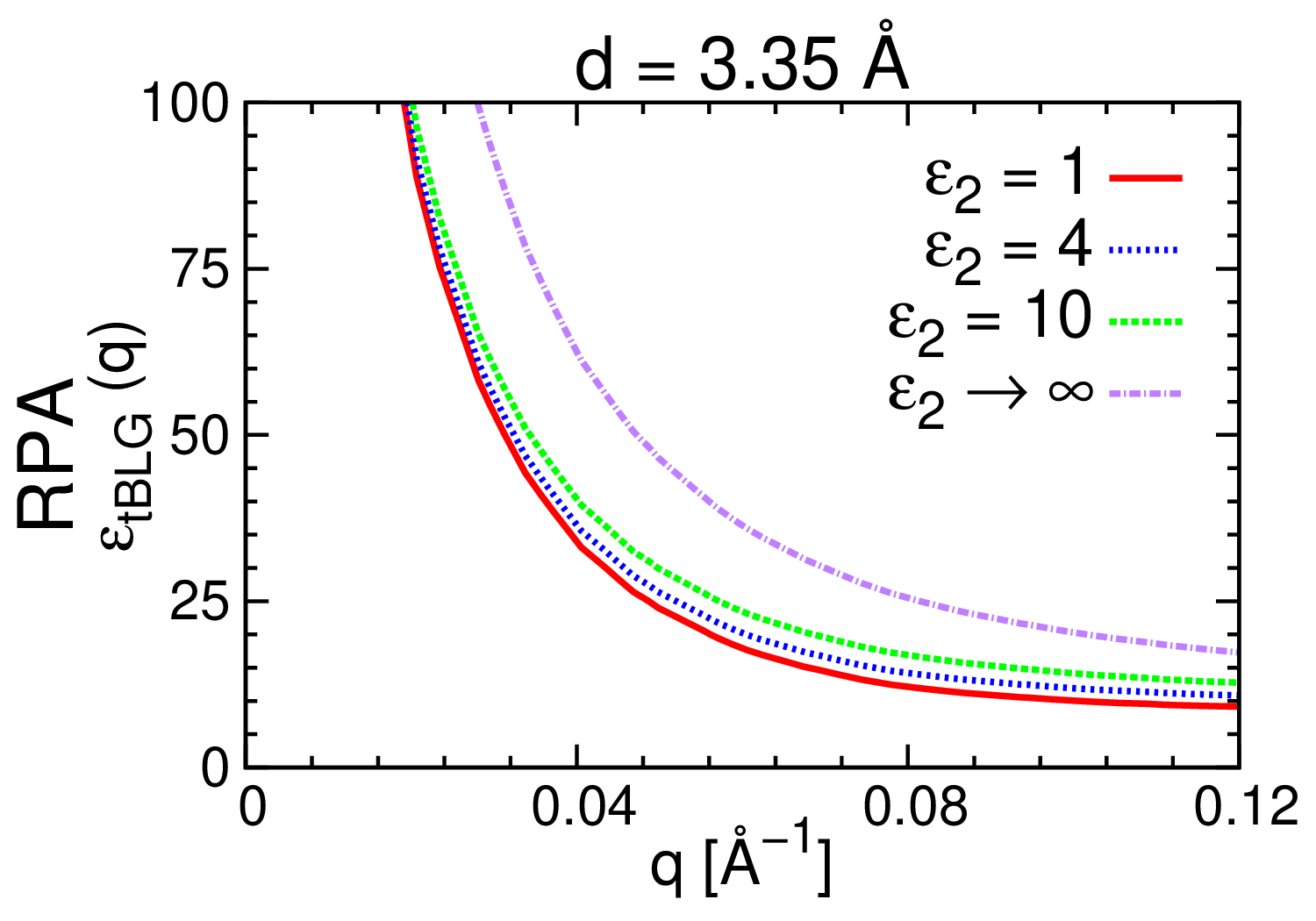}
\includegraphics[clip,width=0.325\textwidth]{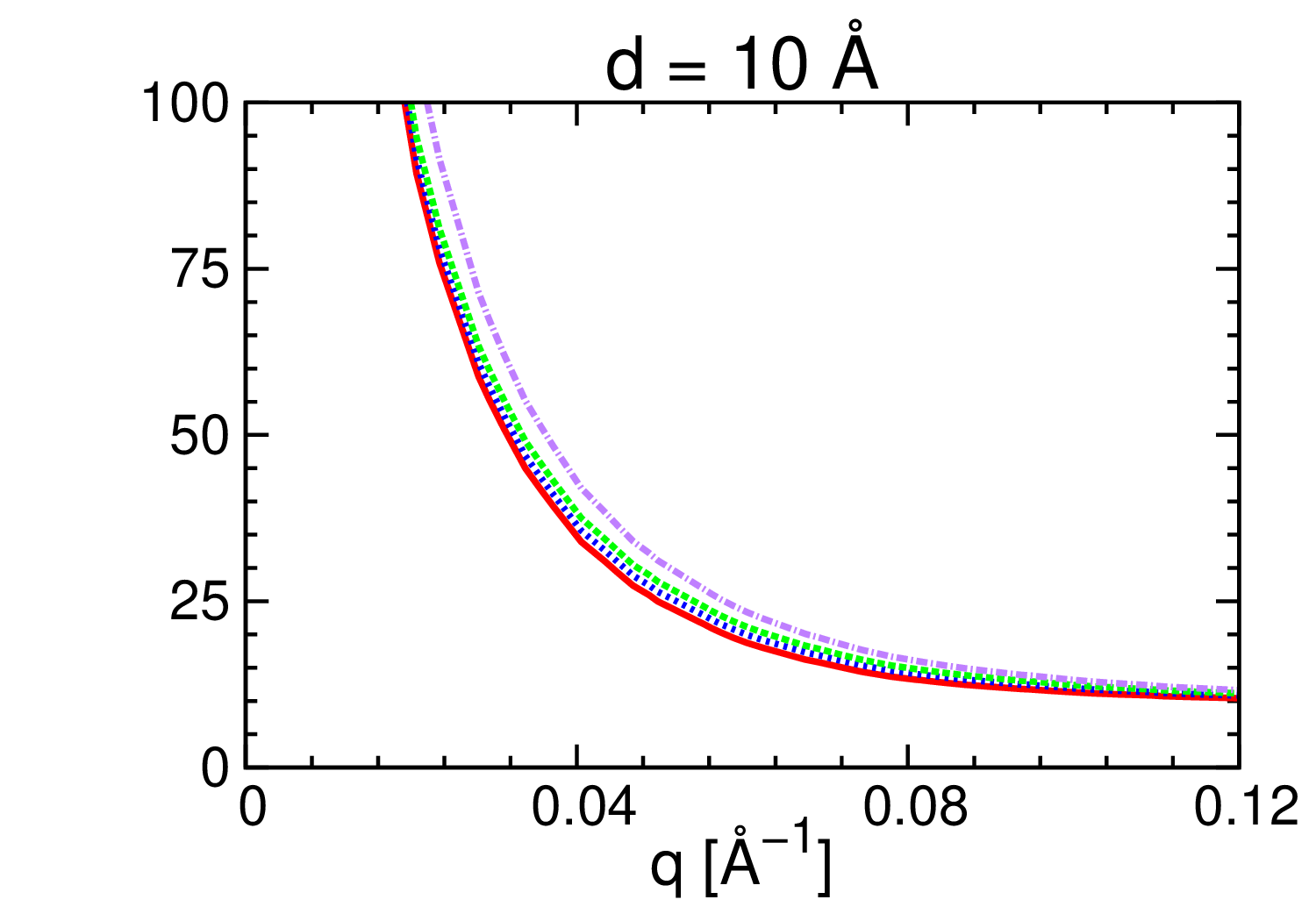} 
\includegraphics[clip,width=0.325\textwidth]{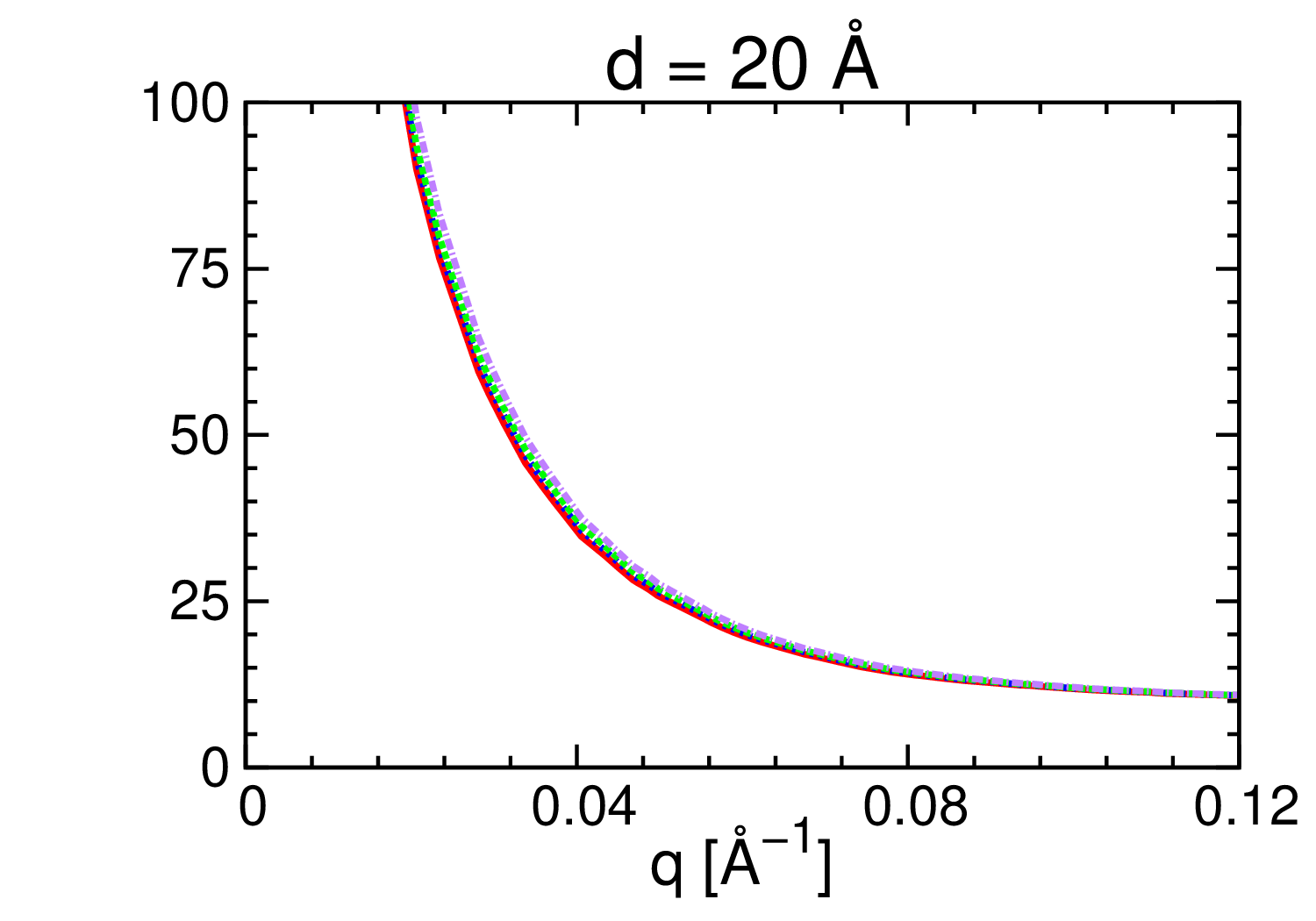} 
\caption{(Color online) cRPA (top row) and RPA (bottom row) dielectric functions as function of the scattering wave vector $q$ for various capping layers $\varepsilon_2$ at different distances $d=3.35 \: \text{\AA}$ (left column), $10 \: \text{\AA}$ (middle column) and $20 \: \text{\AA}$ (right column).} 
\label{figSM3} 
\vspace{-0.3cm}
\end{figure*}

The screening of Coulomb interactions is encoded in the dielectric function $\varepsilon (q) = \varepsilon_{\text{env}}(q) + V(q) \Pi (q)$, where $\varepsilon_{\text{env}} (q)$ as given in Eq. (6) of the main text accounts for the different dielectric environments  (c.f. Fig. 1 of the main text). We neglected local field effects in the study of the dielectric function which is well justified at all $q$ in the cRPA case and at small $q$, which are most relevant here, also in the RPA case (c.f. section 2 of the supplement). Fig. \ref{figSM3} shows the dielectric functions for the cRPA (top row) and RPA (bottom row) for MA-tBLG in different dielectric environments $\varepsilon_2$ at different distances $d=3.35$, $10$ and $20 \: \text{\AA}$.

In the RPA case, the intrinsic screening is strong. Hence, the dielectric environment has a small effect, as seen in Fig. \ref{figSM3}. In the cRPA case, the dielectric environment affects the screening in a considerable $q$-range covering e.g. the whole first mini BZ if the dielectric / metallic gate is as close as $d\lesssim 10 \: \text{\AA}$. These behaviors connect with Fig. 3 of the main text, where we have shown that the effective local interaction remains almost unaffected by the dielectric surrounding in the metallic state (RPA case) but can vary up to $40-50 \%$ in the Mott insulating state (cRPA case).





\subsection{TEMPERATURE DEPENDENCE OF THE POLARIZATION FUNCTION}

Fig. \ref{figSM4} shows the temperature dependence of the polarization function $\Pi_0(q)$ as calculated in the RPA and cRPA according to Eq. (1) of the main text. We compare $T\approx 50$\,K (inverse temperature $\beta=200$\,eV$^{-1}$) and $T\approx 10$\,K ($\beta=1000$\,eV$^{-1}$). In the cRPA, $\Pi_0 (q)$ is essentially temperature independent in this range due to the fact that cRPA models an effectively gapped system. In contrast, the RPA polarizability is strongly temperature dependent: in RPA, $\Pi_0(q)$ increases as the temperature is lowered, which is understandable from the band width of the low energy flat bands being on the order of only a few meV. These results confirm the conclusions drawn in the main text: in cRPA, the system is sensitive to the dielectric environment essentially independently of the temperature. In RPA, on the other hand, the system is not affected by the surrounding dielectrics due to strong internal screening. This statement holds already at $T=50$\,K as discussed in the main text and becomes even more pronounced at smaller temperatures as can be seen in Fig. \ref{figSM4}(a).


\begin{figure*}
\leavevmode
\includegraphics[clip,width=0.49\textwidth]{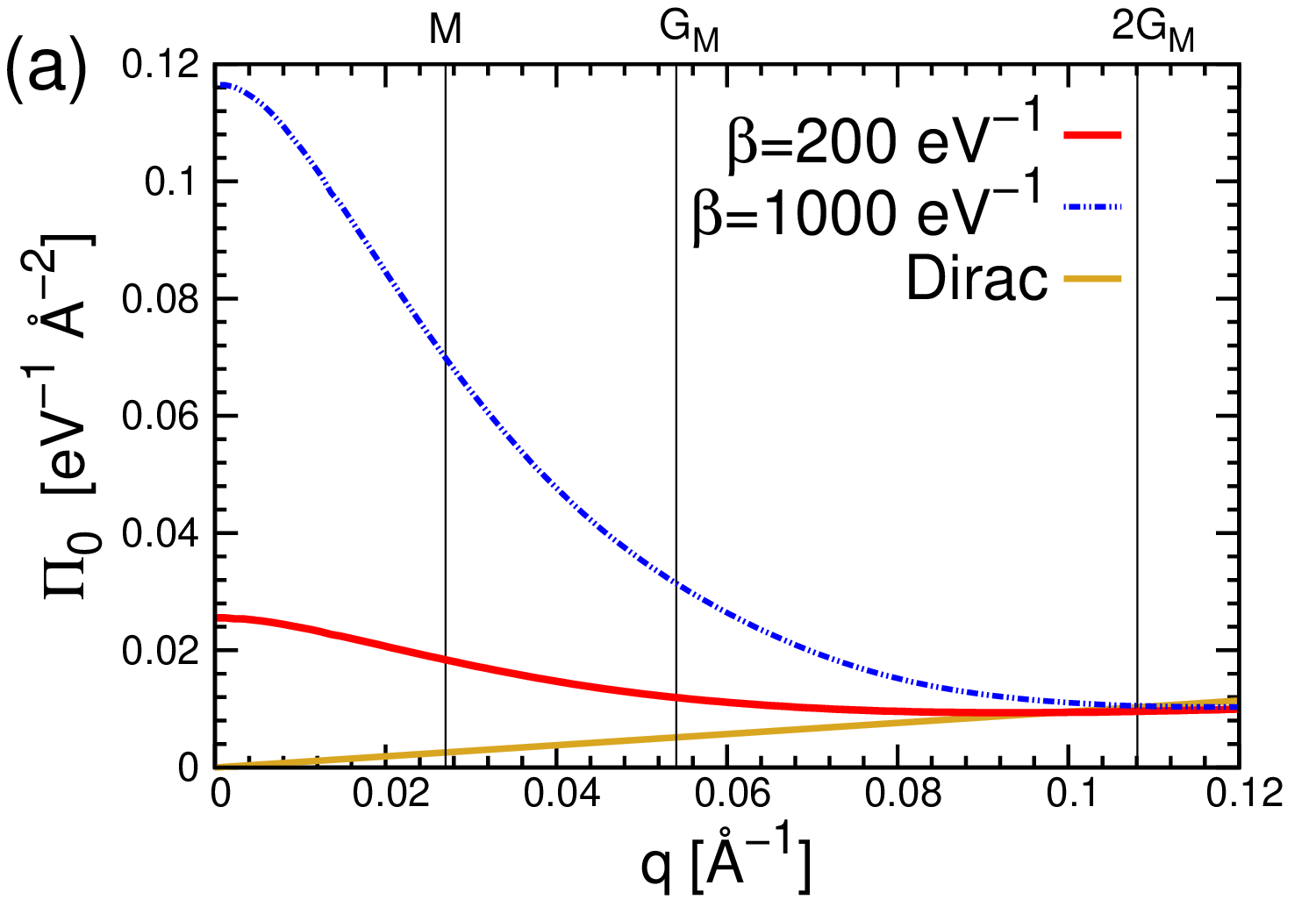}
\includegraphics[clip,width=0.49\textwidth]{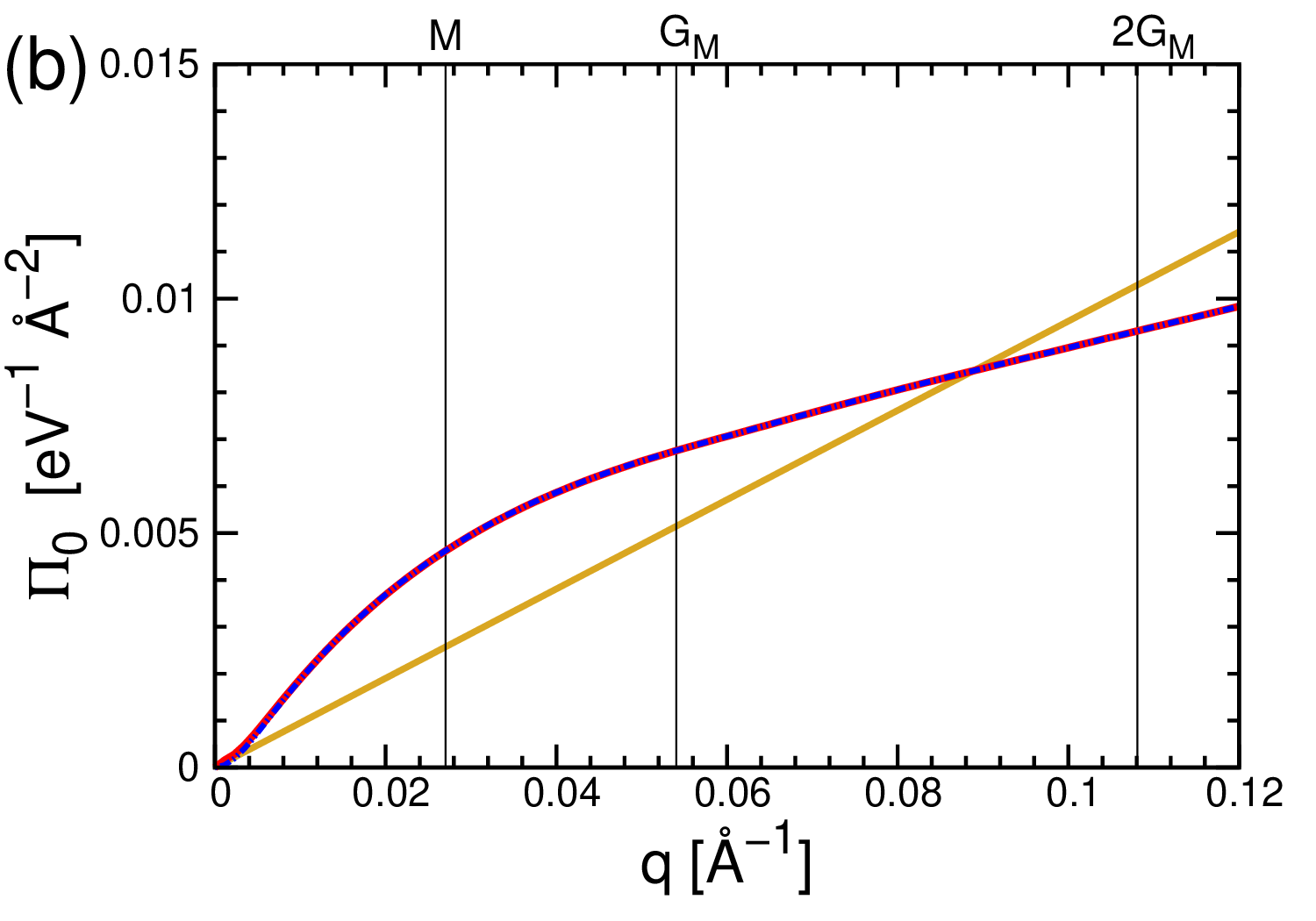}
\caption{(Color online) (a) RPA and (b) cRPA polarization functions as a function of the scattering wave vector $q$ for two different temperatures $\beta=200 \: \text{eV}^{-1}$ ($T \approx 50 \: \text{K}$, red line) and $\beta=1000 \: \text{eV}^{-1}$ ($T \approx 10 \: \text{K}$, blue dash-dotted line) in comparison to uncoupled tBLG ($\Pi_0^{\rm{Dirac}}$, orange line) from Eq. (3) of the main text. $G_c$ is set to 4. In the cRPA case, the polarization function is essentially temperature independent and the curves for $\beta=200 \: \text{eV}^{-1}$ and $\beta=1000 \: \text{eV}^{-1}$ are on top of each other.} 
\label{figSM4} 
\vspace{-0.3cm}
\end{figure*}


\subsection{DEPENDENCE OF INTERNAL SCREENING ON TWIST ANGLES, DOPING, AND INTERLAYER COUPLING PARAMETERS}

We assess in the following the robustness of the conclusions drawn in the main text with respect to variations in the experimental conditions (twist angle and doping) and details of the model employed (interlayer coupling parameters). We study the polarization function of tBLG for twist angles away from the first magic angle $1.05 \: ^{\circ}$, finite doping, and different ratios of AA to AB interlayer coupling $u/u'$, which corresponds to different assumptions on vertical relaxations \cite{BisPNAS1082011,KosPRX82018,tarnopolsky_origin_2019}.


\subsubsection{Twist angle dependence}

\begin{figure*}
\leavevmode
\includegraphics[clip,width=0.48\textwidth]{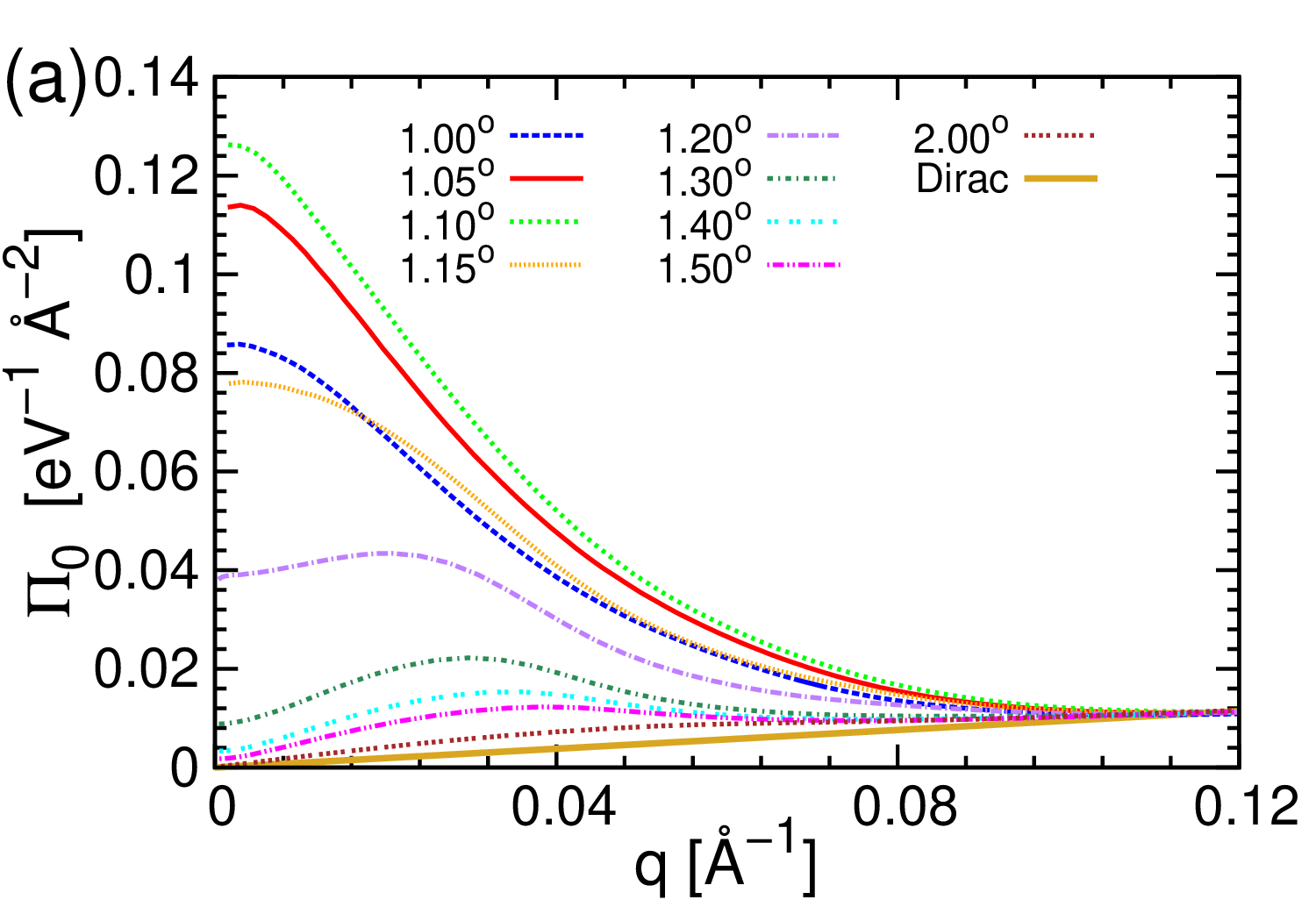}
\includegraphics[clip,width=0.48\textwidth]{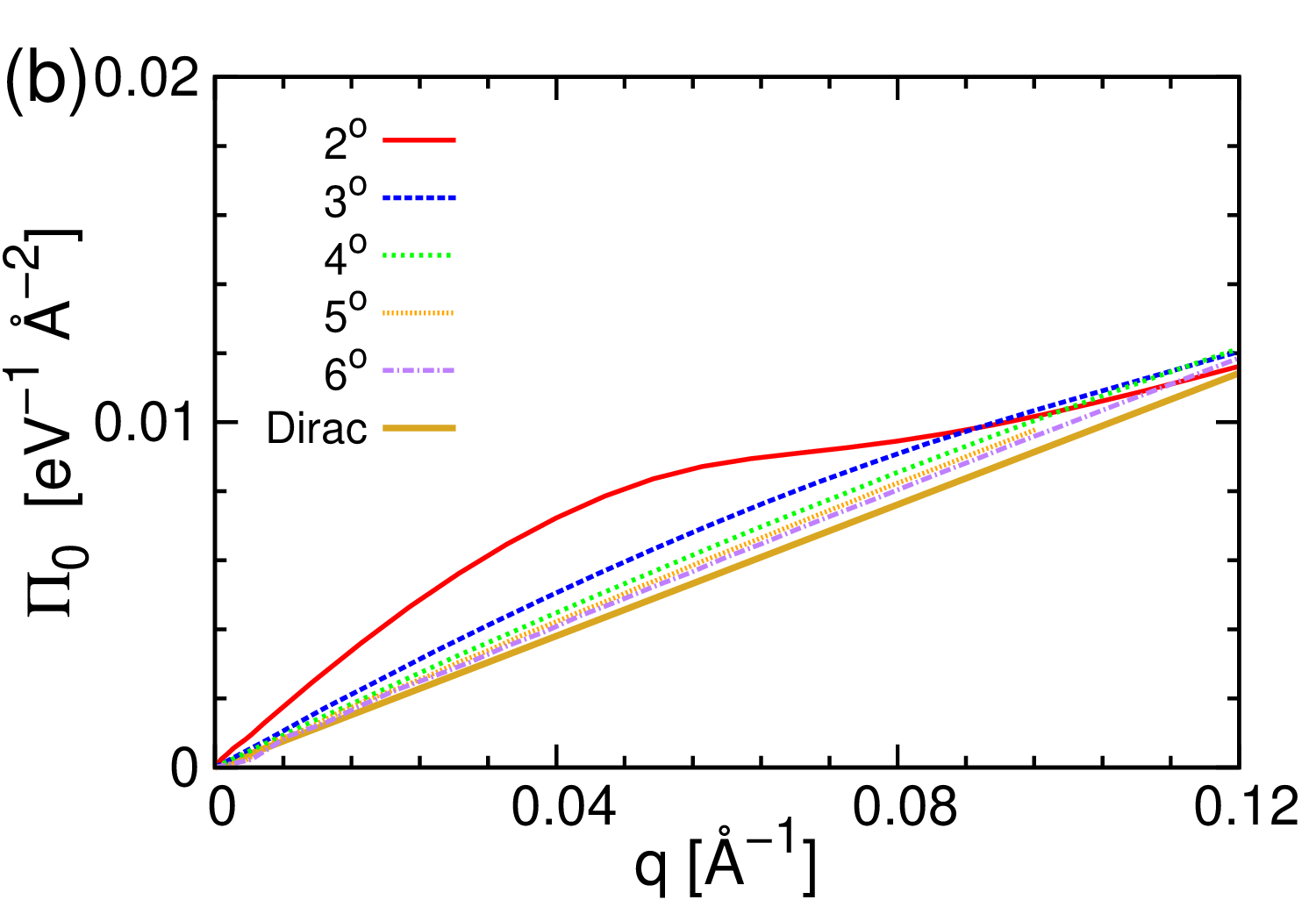}
\includegraphics[clip,width=0.48\textwidth]{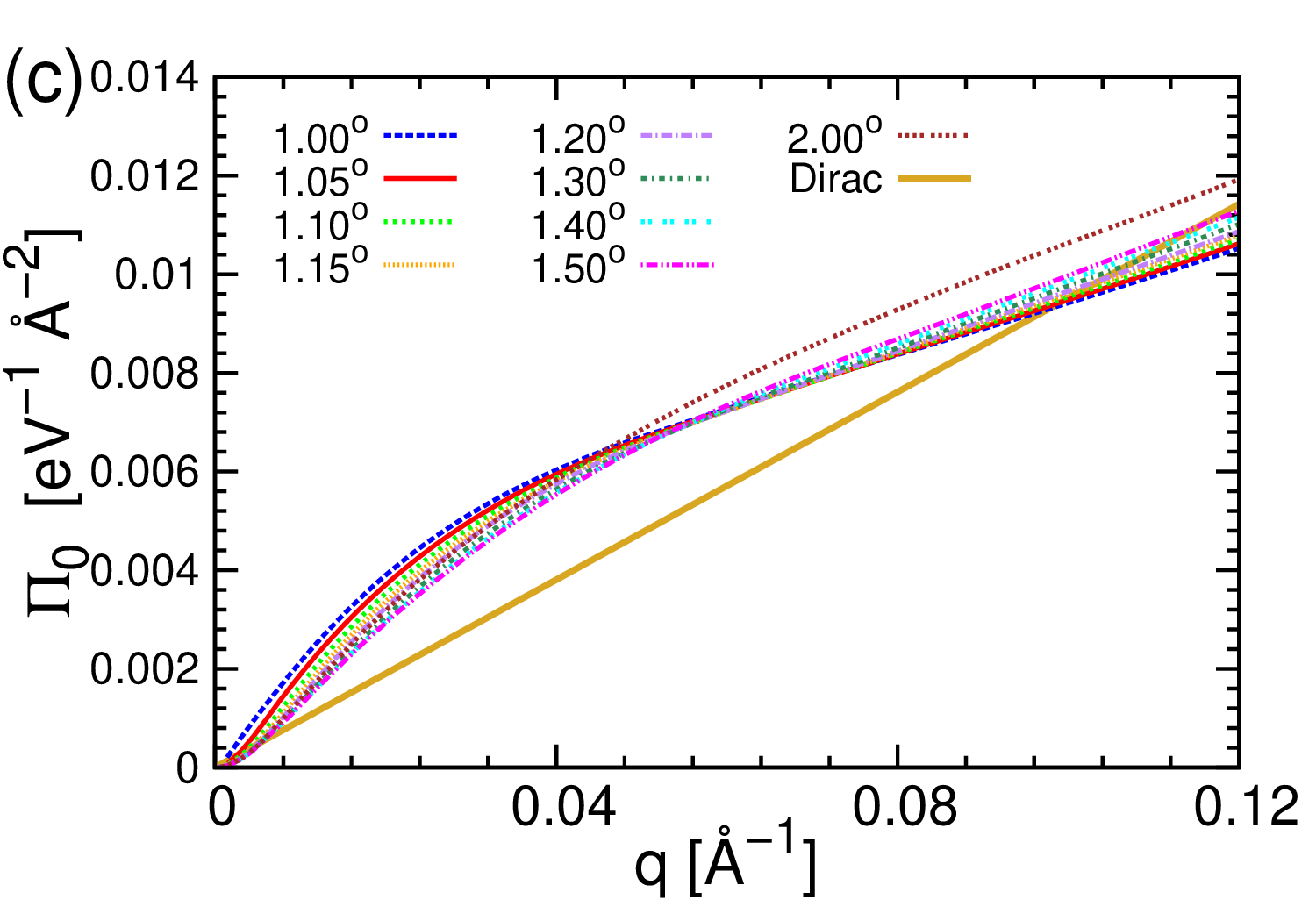}
\includegraphics[clip,width=0.48\textwidth]{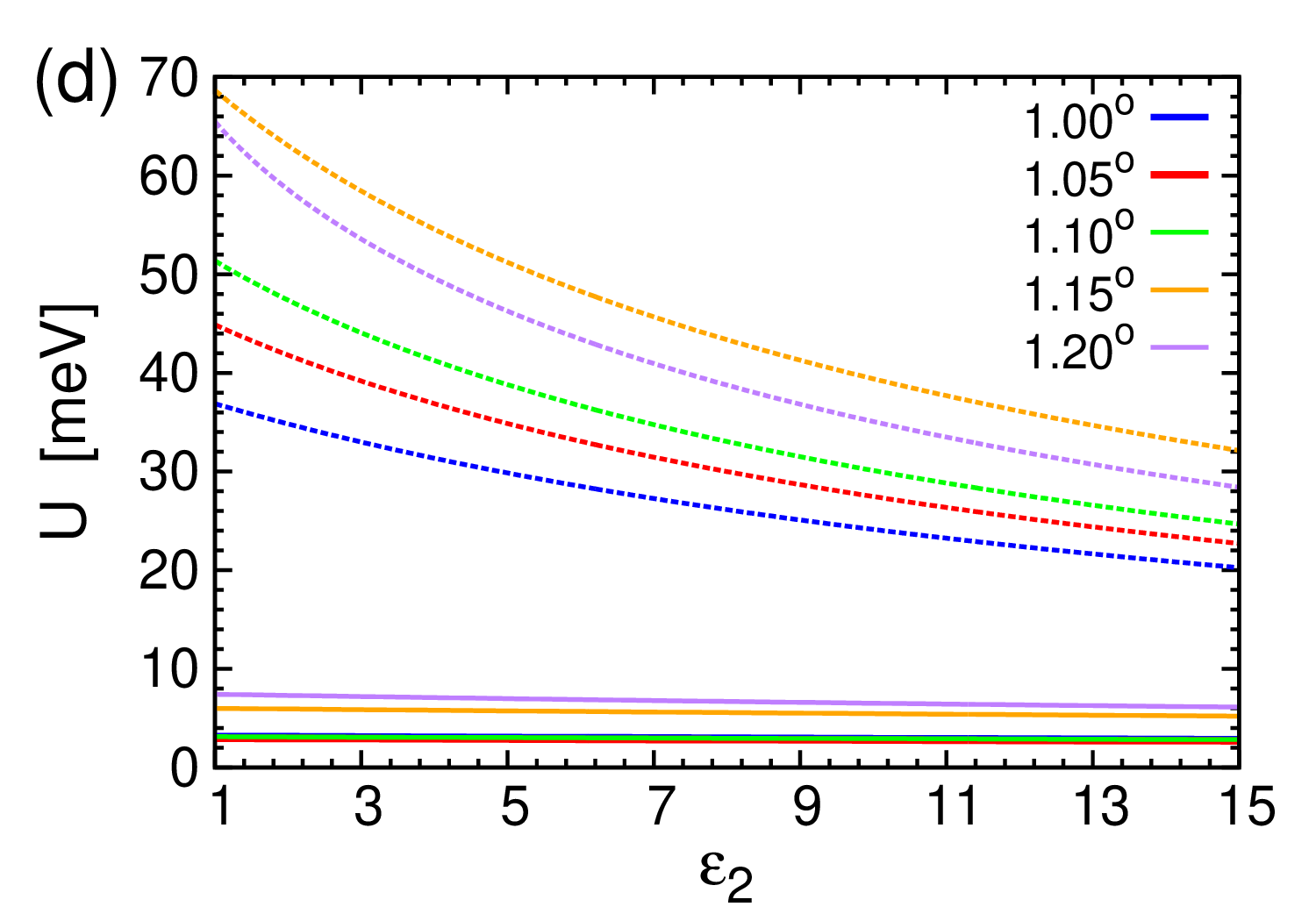}
\caption{(Color online) Twist angle dependence of the  RPA (a, b) and cRPA polarization functions (c) in comparison to the model of uncoupled tBLG ($\Pi_0^{\rm{Dirac}}$, orange continuous line) from Eq. (3) of the main text. Polarization functions calculated for different twist angles between $1.00\: ^{\circ}$ and $6.00\: ^{\circ}$ are shown in different colors. Calculations were carried out at inverse temperature $\beta=1000 \: \text{eV}^{-1}$, and $G_c=6$. (d) Dielectric engineering of tBLG for twist angles $1.00 \: ^{\circ}\leq \theta \leq 1.20 \: ^{\circ}$. The effective interacition as obtained from Eq. (7) of the main text within RPA (solid lines) and cRPA (dotted) is shown. RPA (cRPA) effective interaction is insensitive (senstive) to the dielectric environment in the whole range of twist angles, similarly to the case of the magic angle $\theta=1.05 \: ^{\circ}$ discussed in the main text.} 
\label{figSM5} 
\vspace{-0.3cm}
\end{figure*}

The twist angle dependence of the internal screening in tBLG can be understood from Fig. \ref{figSM5}. The polarization functions obtained in RPA are shown in Fig. \ref{figSM5}(a) for twist angles $1.00 \: ^{\circ}\leq \theta \leq 2.00 \: ^{\circ}$ and in Fig. \ref{figSM5}(b) for $\theta \geq 2.00 \: ^{\circ}$.  Within the range $1.00 \: ^{\circ}\leq \theta \leq 1.20 \: ^{\circ}$ the RPA polarization functions behave qualitatively similar to the magic angle case of $\theta=1.05 \: ^{\circ}$ in the sense that the RPA polarization functions exceed by far the model of uncoupled Dirac fermions. For $\theta > 2 \: ^{\circ}$ the situation changes and the RPA screening recovers the linear-$q$ dependence as expected from Eq. (3), but with a renormalized Fermi velocity $v_F^*$, where $v_F^* (\theta=3\: ^{\circ})/v_F \approx 0.882$, $v_F^* (\theta=4\: ^{\circ})/v_F \approx 0.895$, $v_F^* (\theta=5\: ^{\circ})/v_F \approx 0.927$ and $v_F^* (\theta=6\: ^{\circ})/v_F \approx 0.958$. In the cRPA case (Fig. \ref{figSM5}(c)), the low $q$ behavior is similar for all twist angles $\theta<2 \: ^{\circ}$. Thus, the conclusions on the interplay of external versus internal screening drawn in the main text will be valid not only at the magic angle but in an extented range $1.00 \: ^{\circ}\leq \theta \leq 1.20 \: ^{\circ}$ around the magic angle.

In Fig. \ref{figSM5}(d), we show impact of dielectric engineering on the effective local interaction $U$ as obtained in RPA and cRPA for twisting angles $1.00 \: ^{\circ}\leq \theta \leq 1.20 \: ^{\circ}$. We considered a dielectric $\varepsilon_2$ in direct contact with the tBLG corresponding to the setup depicted in Fig. 1(b) of the main text, and calculated $U$ according to Eq. (7) of the main text with $q_c$ being rescaled in all cases according to the twist-angle dependent moiré lattice constants. In the entire range of twist angles considered, here, the RPA effective interactions $U$ are insensitive to the dielectric environment, while the effective interactions in cRPA change with $\varepsilon_2$ in a similar manner for all twisting angles. Therefore, we can conclude that our main results are robust against variations of the twist angle around the magic angle in the range $1.00 \: ^{\circ}\leq \theta \leq 1.20 \: ^{\circ}$. This range of twist angles is far larger than uncertainties of twist angles in experiments of Refs. \onlinecite{CaoN5562018ins,CaoN5562018sc,lu_superconductors_2019} and covers the range of angles within which possible unconventional superconducting and insulating states have been observed therein. 

\subsubsection{Doping dependence}

\begin{figure*}
\leavevmode
\includegraphics[clip,width=0.48\textwidth]{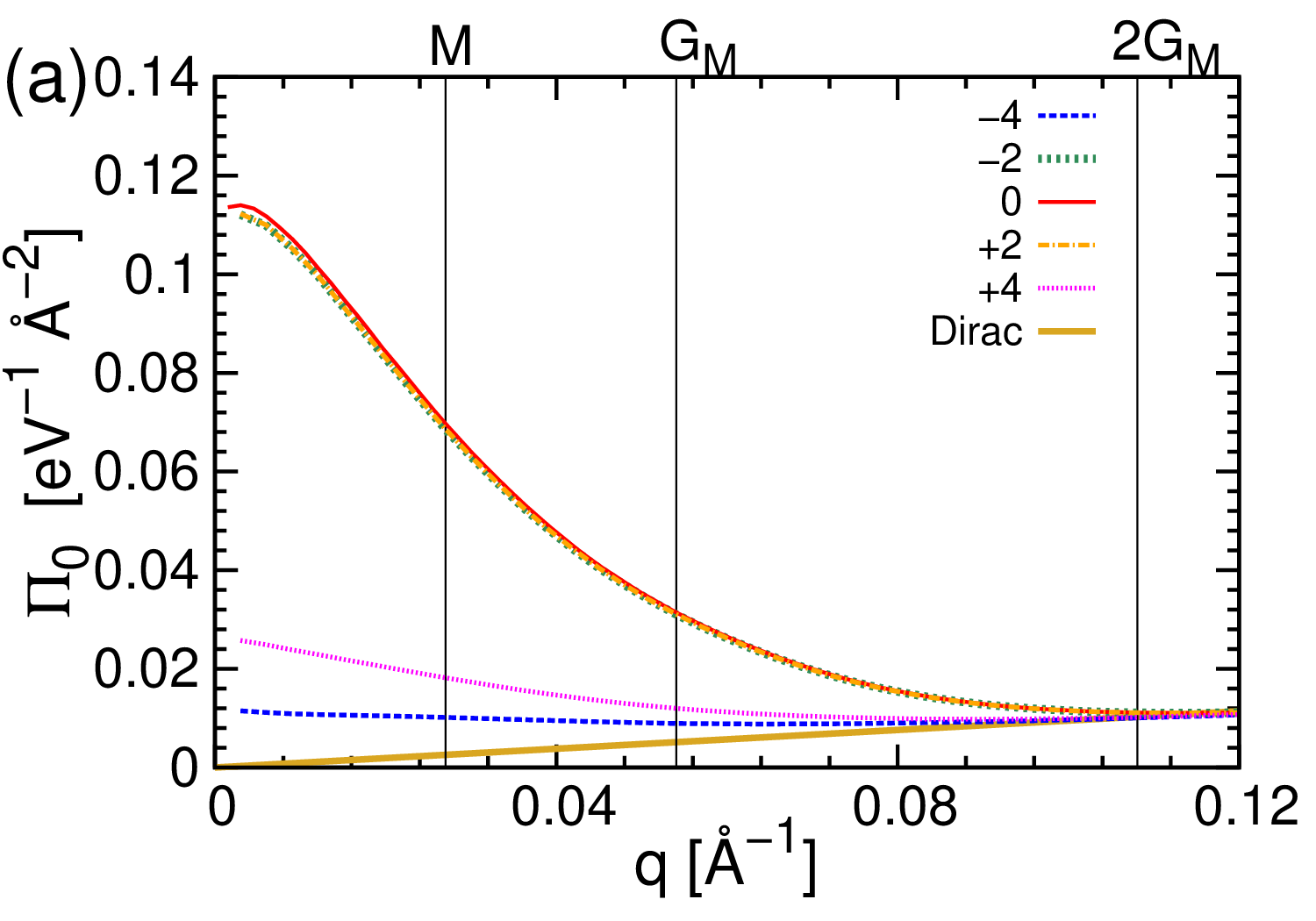}
\includegraphics[clip,width=0.48\textwidth]{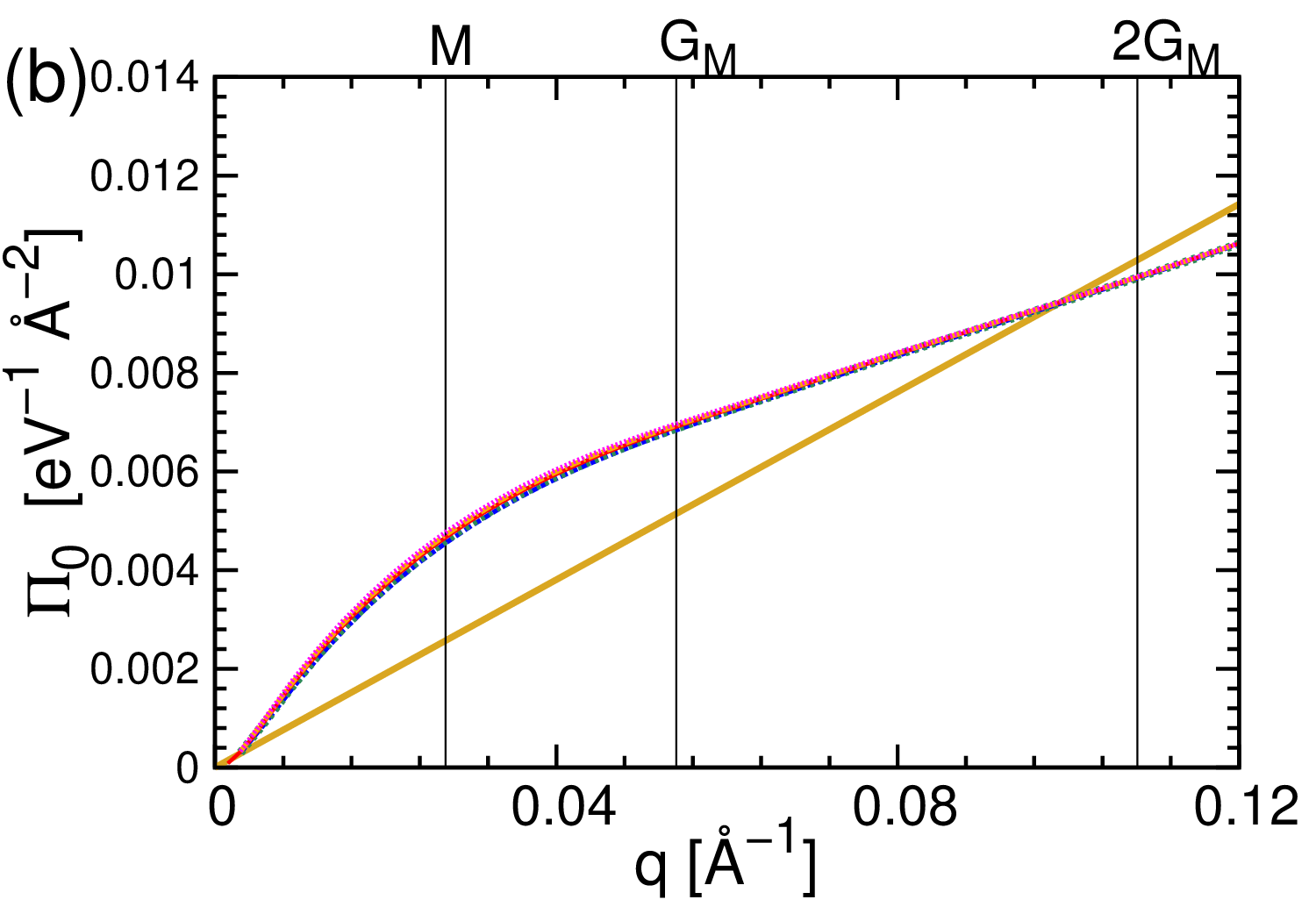}
\caption{(Color online) Doping dependence of the RPA (a) and cRPA polarization functions (b) in comparison to the model of charge neutral uncoupled tBLG ($\Pi_0^{\rm{Dirac}}$, orange continuous line) from Eq. (3) of the main text. Calculations were performed at inverse temperature $\beta=1000 \: \text{eV}^{-1}$, and $G_c=6$. Polarization functions for different moiré band filling factors $-4\leq\nu\leq 4$ are marked in different colors. $\nu=0$ corresponds to the CNP, $\nu = \pm 2$ to half-filling of the flat bands, and $\nu = \pm 4$ describes fully empty / filled flat bands. 
} 
\label{figSM6} 
\vspace{-0.3cm}
\end{figure*}

In Fig. \ref{figSM6}, we show RPA/cRPA polarization functions for different levels of doping. The band filling $\nu=n/n_0$ is given relative to the CNP $\nu = 0$ in multiples of the density $n_0$ needed to fill \textit{one} moir\'e superlattice band. Due to spin and valley degeneracy, $\nu = \pm 4$ then describes fully empty / filled flat bands and $\nu = \pm 2$ the half-filled flat bands. Around half-filling of the positive and negative flat bands (i.e. $\nu = \pm 2$), we find that RPA screening is of the same order as in the undoped case. Significant changes occur in the RPA case only in the limit of completely empty / filled flat bands ($\nu \approx \pm 4$). The cRPA results are virtually independent of the doping level. Thus, our results hold for the range of dopings at which unexpected superconducting and insulating states have been observed in Refs. \onlinecite{CaoN5562018ins,CaoN5562018sc,lu_superconductors_2019}.

\subsubsection{Interlayer coupling dependence}

\begin{figure*}
\leavevmode
\includegraphics[clip,width=0.48\textwidth]{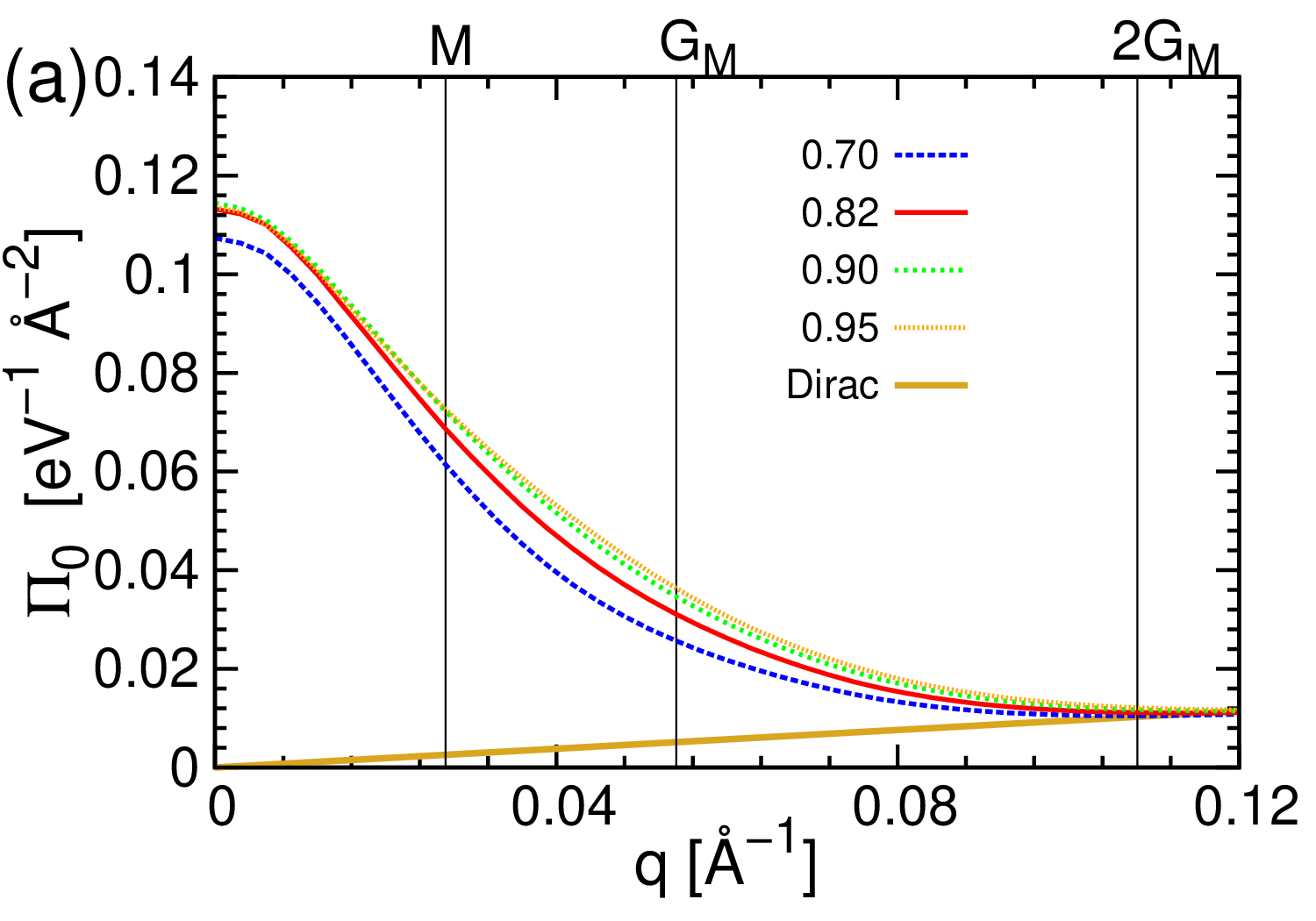}
\includegraphics[clip,width=0.48\textwidth]{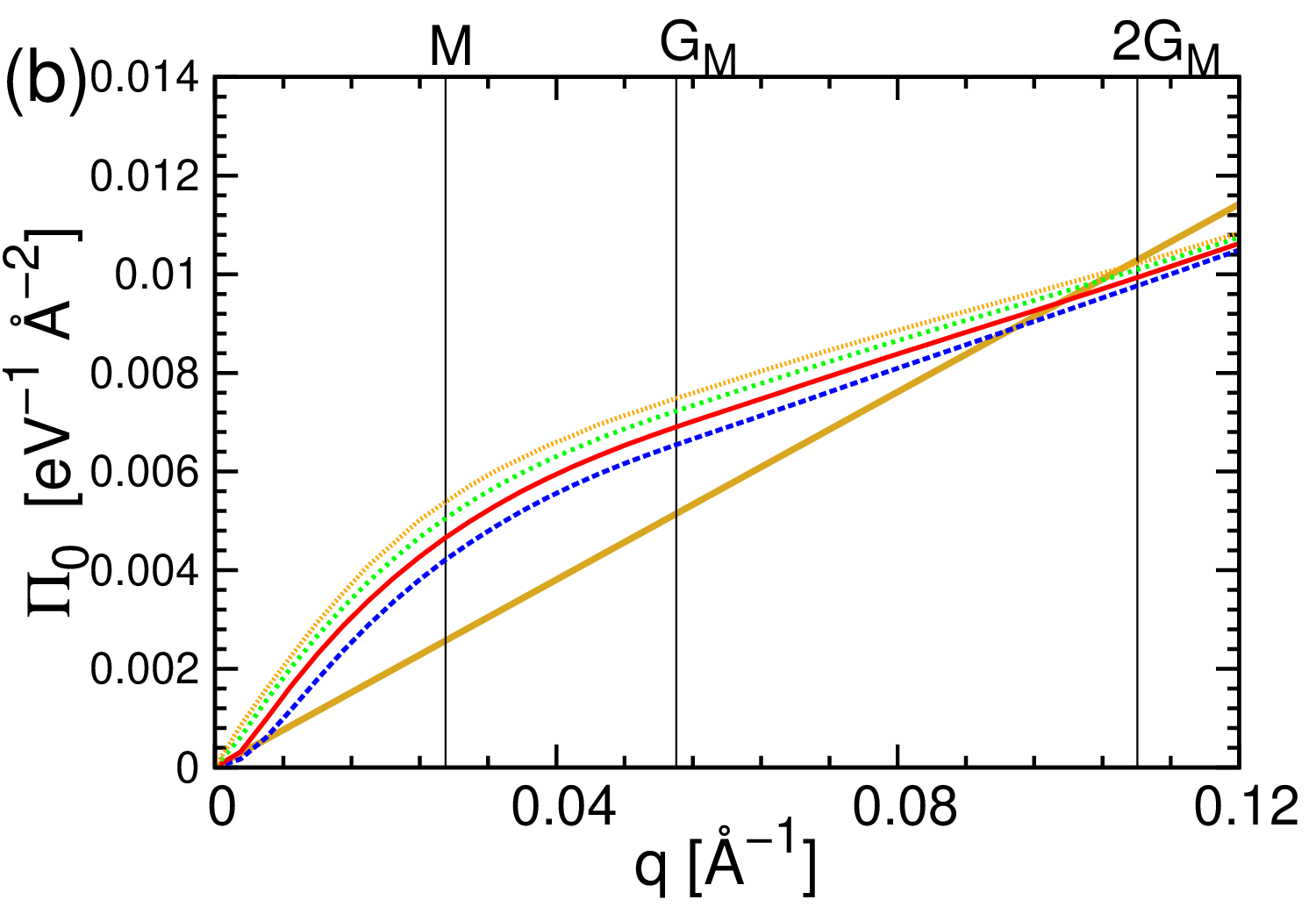}
\caption{(Color online) Dependence of the RPA (a) and cRPA polarization functions (b) on the ratio of AA to AB interlayer coupling $u/u'$. The model of uncoupled tBLG ($\Pi_0^{\rm{Dirac}}$, orange continuous line) from Eq. (3) of the main text is shown for comparison. Inverse temperature $\beta=1000 \: \text{eV}^{-1}$ and $G_c=6$. $u/u'=0.82$ refers to the model considered in the main text and in \cite{KosPRX82018}. $u/u' = 1$ describes the non-corrugated case assumed in \cite{BisPNAS1082011}.} 
\label{figSM7} 
\vspace{-0.3cm}
\end{figure*}

The continuum model employed, here, depends sensitively on the ratios of AA to AB interlayer coupling $u/u'$, which encodes vertical relaxation effects. Different $u/u'$ ratios have been employed in the literature \cite{BisPNAS1082011,KosPRX82018,tarnopolsky_origin_2019}. The work of Ref. \onlinecite{BisPNAS1082011} assumed equal AA and AB interlayer coupling $u/u'=1$, which neglects the larger interlayer distance in the AA regions as compared to the AB regions. To account for these relaxation effects, $u/u' \approx 0.82$ has been considered in Ref. \onlinecite{KosPRX82018} and also used for all calculations shown in the main text of this work. In the following, we study the impact of variations in the $u/u'$ ratio on our results: Fig \ref{figSM7} shows RPA/cRPA polarization functions for MA-tBLG obtained with $u/u' = 0.70$, $0.82$ \cite{KosPRX82018}, $0.90$ and $0.95$ . We obtain almost the same polarization functions as obtained from the model considered in the main text ($u/u'=0.82$ \cite{KosPRX82018}). Thus, our conclusions are also robust against variations of the $u/u'$ ratio in the low-energy continuum model.

Taken together, we have shown that the conclusions drawn on the interplay of internal and external screening in MA-tBLG are robust against variations of system parameters like twist angle, doping, and details of the model parametrization.

\subsection{INFLUENCE OF ENERGY GAPS ON INTERNAL SCREENING OF MA-tBLG}

In the experiments of Refs. \cite{CaoN5562018ins,CaoN5562018sc,Yankowitz1059,lu_superconductors_2019} the opening of energy gaps at integer filling factors $\nu$ has been reported. In particular, gaps on the order of $0.1$~meV - $0.3$~meV at half-filling and other integer fillings $\nu\neq 0$ were found experimentally in Refs. \cite{CaoN5562018ins,CaoN5562018sc,Yankowitz1059,lu_superconductors_2019}. Gaps found at the CNP ($\nu=0$) are likely larger \cite{2019arXiv190102997C,lu_superconductors_2019}. Currently, the nature of all these gaps is unclear. If (some of) these gaps arise from a weak coupling scenario, it can be expected that screening processes from a low energy window on the order of the gap are suppressed, while all higher energy contributions to the polarization function remain essentially unaffected. To mimic such a situation, we study how exclusion of polarization processes taking place inside an energy window of $\pm \Delta$ around the chemical potential affects the polarization function. To this end, we modify Eq. (1) from the main text to calculate the polarization function in the following manner:
\begin{widetext}
\begin{equation}
\Pi_0^{\bold{G}, \bold{G}'} (\bold{q}) = \frac{g_s g_v}{S_M N} \sum_{\substack{\bold{k} \\ \alpha , \beta \\ \bold{G}_2, \bold{G}'_2}} \mathcal{M}_{\bold{G}_2, \bold{G}'_2, \bold{G}, \bold{G}'}^{\alpha \beta} \frac{f_\bold{k}^\alpha - f_{\bold{k}+\bold{q}}^{\beta}}{i\eta + E_\bold{k}^\alpha - E_{\bold{k}+\bold{q}}^\beta} [1-\Theta(\Delta-|E_{\bold{k}+\bold{q}}^\beta-\mu|)\Theta(\Delta-|E_\bold{k}^\alpha-\mu|)].
\label{eq:cRPA_mod}
\end{equation}
\end{widetext}
This formula reduces to Eq. (1) from the main text for $\Delta\to 0$ and corresponds to a flavor of cRPA, where all states from a certain energy window $\pm \Delta$ are assumed to form the correlated subspace. In this flavor of RPA, $f_\bold{k}^\alpha$ denote the non-modified Fermi functions belonging to band $\alpha$ at crystal momentum $\bold{k}$. 

In Fig. \ref{figSM8}, we show the polarization function of MA-tBLG calculated according to Eq. (\ref{eq:cRPA_mod}) with gaps in the range of $\Delta = 0.05 \: \text{meV}$ to $1 \: \text{meV}$ being included in the simulations. For $\Delta \leq 0.15 \: \text{meV}$, the polarization functions are very similar to the ones obtained within full RPA. Thus, weak coupling instabilities opening gaps in the $\lesssim 0.3 \: \text{meV}$ range can be expected to barely affect the internal polarization function of tBLG. Correspondingly, we expect that the RPA results shown in the main text are representative for conventional metallic states and gapped states resulting from weak coupling mechanism, as assumed in the main text. For gaps opened by strong coupling mechanisms, the situation can be very different as explained in the main text and illustrated in the $\Delta=1 \: \text{meV}$ case in Fig. \ref{figSM8}.

\begin{figure*}
\leavevmode
\includegraphics[clip,width=0.325\textwidth]{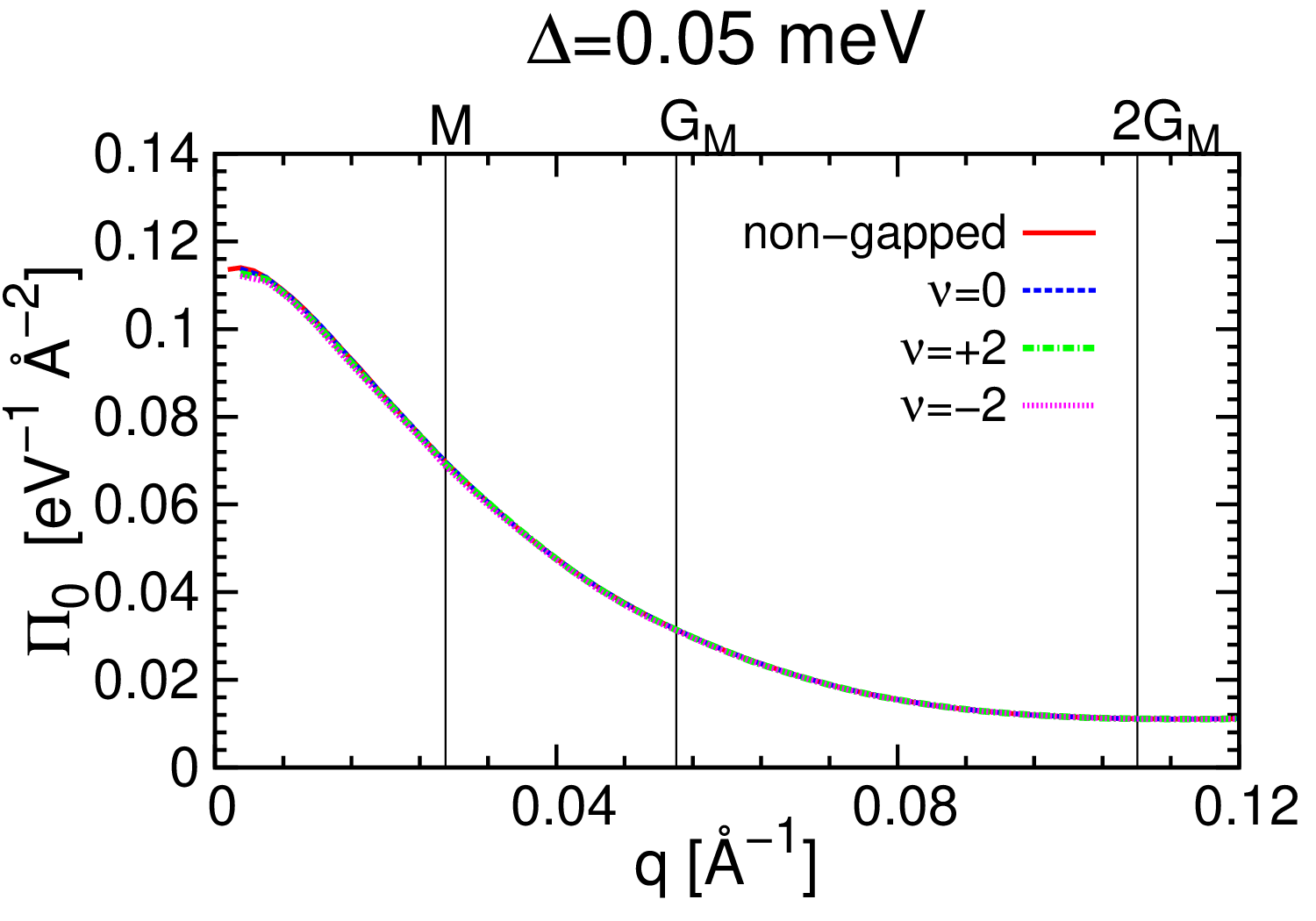}
\includegraphics[clip,width=0.325\textwidth]{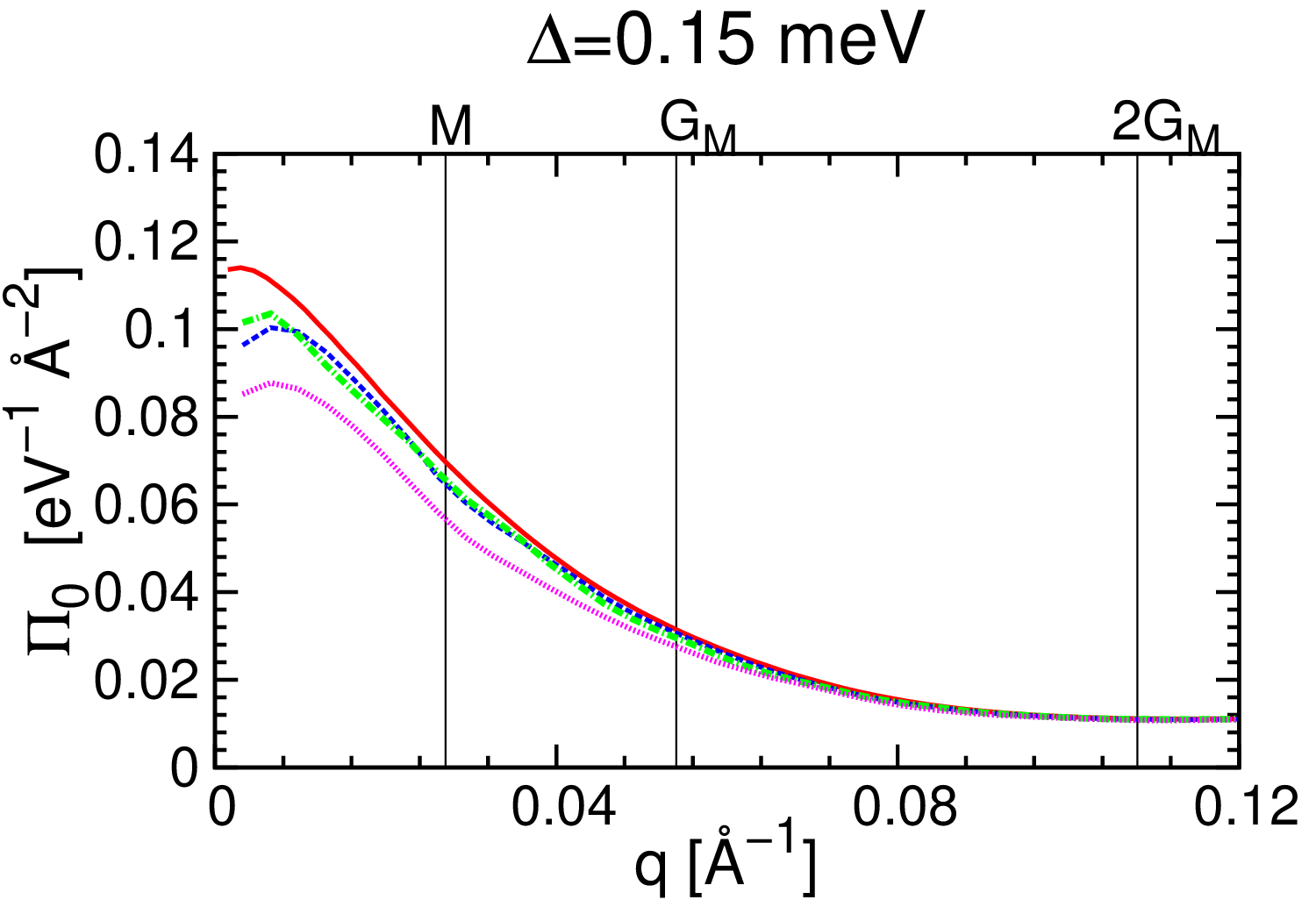}
\includegraphics[clip,width=0.325\textwidth]{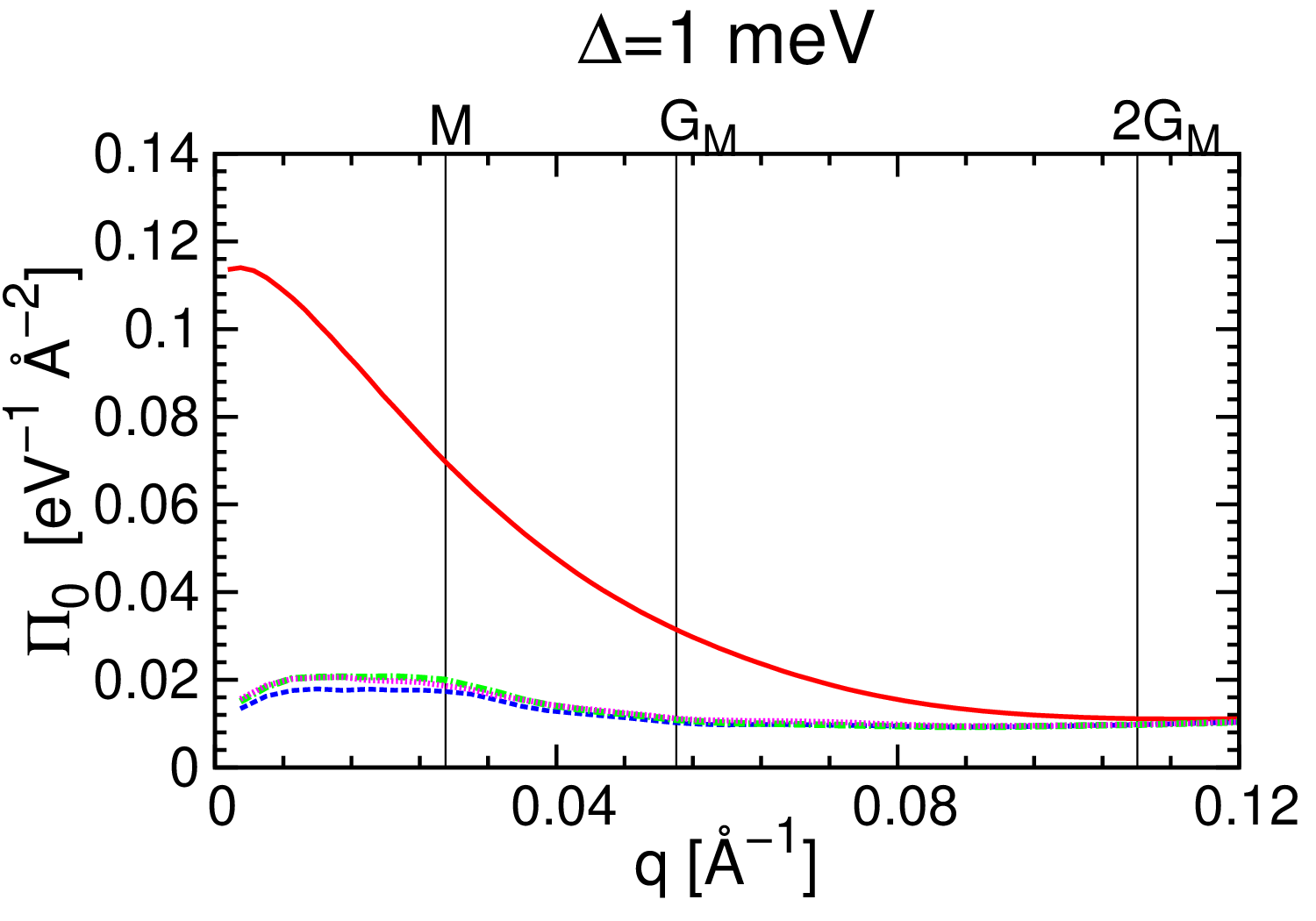}
\caption{(Color online) Polarization functions of MA-tBLG excluding of polarization processes taking place inside an energy windows of $\pm \Delta$ around the chemical potential according to Eq. (\ref{eq:cRPA_mod}). For each energy window / effective gap size $\Delta$ three fillings are considere $\nu=\pm 2$ (half-filling) and $\nu= 0$ (charge neutraility). Calculations were performed at inverse temperature $\beta=1000 \: \text{eV}^{-1}$, and $G_c=6$.} 
\label{figSM8} 
\vspace{-0.3cm}
\end{figure*}

}

\end{document}